\documentclass[aps,prb,reprint,noshowkeys,superscriptaddress]{revtex4-1}
\usepackage{graphicx,dcolumn,bm,xcolor,microtype,multirow,amscd,amsmath,amssymb,amsfonts,physics,longtable,wrapfig,txfonts}
\usepackage[version=4]{mhchem}

\usepackage{natbib}
\usepackage[extra]{tipa}
\AtBeginDocument{\nocite{achemso-control}}

\usepackage[utf8]{inputenc}
\usepackage[T1]{fontenc}
\usepackage{txfonts}

\usepackage[
	colorlinks=true,
    citecolor=blue,
    breaklinks=true
	]{hyperref}
\newcommand{\ie}{\textit{i.e.}}
\newcommand{\eg}{\textit{e.g.}}

\definecolor{darkgreen}{HTML}{009900}
\usepackage[normalem]{ulem}
\newcommand{\titou}[1]{\textcolor{black}{#1}}

\newcommand{\mc}{\multicolumn}
\newcommand{\fnm}{\footnotemark}
\newcommand{\fnt}{\footnotetext}
\newcommand{\tabc}[1]{\multicolumn{1}{c}{#1}}
\newcommand{\SI}{\textcolor{blue}{supporting information}}

\newcommand{\T}[1]{#1^{\intercal}}

\newcommand{\br}[1]{\mathbf{r}_{#1}}
\newcommand{\dbr}[1]{d\br{#1}}

\newcommand{\evGW}{ev$GW$}	
	
\newcommand{\GOWO}{$G_0W_0$}	

\newcommand{\xc}{\text{xc}}
\newcommand{\Ha}{\text{H}}

\newcommand{\Norb}{N}
\newcommand{\Nocc}{O}
\newcommand{\Nvir}{V}
\newcommand{\IS}{\lambda}

\newcommand{\Enuc}{E^\text{nuc}}
\newcommand{\Ec}{E_\text{c}}
\newcommand{\EHF}{E^\text{HF}}
\newcommand{\EBSE}{E^\text{BSE}}

\newcommand{\EcBSE}{E_\text{c}^\text{BSE}}

\newcommand{\Req}{R_\text{eq}}

\newcommand{\eHF}[1]{\epsilon^\text{HF}_{#1}}

\newcommand{\eGW}[1]{\epsilon^{GW}_{#1}}

\newcommand{\Om}[2]{\Omega_{#1}^{#2}}

\newcommand{\tA}[2]{\Tilde{A}_{#1}^{#2}}

\newcommand{\ABSE}[2]{A_{#1}^{#2,\text{BSE}}}
\newcommand{\BBSE}[2]{B_{#1}^{#2,\text{BSE}}}
\newcommand{\ARPA}[2]{A_{#1}^{#2,\text{RPA}}}
\newcommand{\BRPA}[2]{B_{#1}^{#2,\text{RPA}}}
\newcommand{\ARPAx}[2]{A_{#1}^{#2,\text{RPAx}}}
\newcommand{\BRPAx}[2]{B_{#1}^{#2,\text{RPAx}}}
\newcommand{\G}[1]{G_{#1}}
\newcommand{\LBSE}[1]{L_{#1}}
\newcommand{\XiBSE}[1]{\Xi_{#1}}

\newcommand{\W}[2]{W_{#1}^{#2}}

\newcommand{\vc}[1]{v_{#1}}
\newcommand{\Sig}[1]{\Sigma_{#1}}
\newcommand{\SigGW}[1]{\Sigma^{GW}_{#1}}

\newcommand{\MO}[1]{\phi_{#1}}
\newcommand{\ERI}[2]{(#1|#2)}
\newcommand{\sERI}[3]{[#1|#2]^{#3}}

\newcommand{\OmRPA}[2]{\Omega_{#1}^{#2,\text{RPA}}}

\newcommand{\bO}{\mathbf{0}}
\newcommand{\bI}{\mathbf{1}}

\newcommand{\bA}[1]{\mathbf{A}^{#1}}
\newcommand{\btA}[1]{\Tilde{\mathbf{A}}^{#1}}
\newcommand{\bB}[1]{\mathbf{B}^{#1}}
\newcommand{\bX}[1]{\mathbf{X}^{#1}}
\newcommand{\bY}[1]{\mathbf{Y}^{#1}}
\newcommand{\bZ}[2]{\mathbf{Z}_{#1}^{#2}}
\newcommand{\bK}{\mathbf{K}}
\newcommand{\bP}[1]{\mathbf{P}^{#1}}

\newcommand{\NEEL}{Universit\'e Grenoble Alpes, CNRS, Institut NEEL, F-38042 Grenoble, France}
\newcommand{\CEISAM}{Laboratoire CEISAM - UMR CNRS 6230, Universit\'e de Nantes, 2 Rue de la Houssini\`ere, BP 92208, 44322 Nantes Cedex 3, France}
\newcommand{\LCPQ}{Laboratoire de Chimie et Physique Quantiques (UMR 5626), Universit\'e de Toulouse, CNRS, UPS, France}
\newcommand{\CEA}{Universit\'e Grenoble Alpes, CEA, IRIG-MEM-L Sim, 38054 Grenoble, France}

\begin{document}	

\title{Pros and Cons of the Bethe-Salpeter Formalism for Ground-State Energies}

\author{Pierre-Fran\c{c}ois \surname{Loos}}
	\email{loos@irsamc.ups-tlse.fr}
	\affiliation{\LCPQ}
\author{Anthony \surname{Scemama}}
	\email{scemama@irsamc.ups-tlse.fr}
	\affiliation{\LCPQ}
\author{Ivan \surname{Duchemin}}
	\email{ivan.duchemin@cea.fr}
	\affiliation{\CEA}
\author{Denis \surname{Jacquemin}}
	\email{denis.jacquemin@univ-nantes.fr}
	\affiliation{\CEISAM}
\author{Xavier \surname{Blase}}
	\email{xavier.blase@neel.cnrs.fr }
	\affiliation{\NEEL}

\begin{abstract}
The combination of the many-body Green's function $GW$ approximation and the Bethe-Salpeter equation (BSE) formalism has shown to be a promising alternative to time-dependent density-functional theory (TD-DFT) for computing vertical transition energies and oscillator strengths in molecular systems. 
The BSE formalism can also be employed to compute ground-state correlation energies thanks to the adiabatic-connection fluctuation-dissipation theorem (ACFDT).
Here, we study the topology of the ground-state potential energy surfaces (PES) of several diatomic molecules near their equilibrium bond length. 
Thanks to comparisons with state-of-art computational approaches (CC3), we show that ACFDT@BSE is surprisingly accurate, and can even compete with lower-order coupled cluster methods (CC2 and CCSD) in terms of total energies and equilibrium bond distances for the considered systems.  
However, we sometimes observe unphysical irregularities on the ground-state PES in relation with difficulties in the identification of a few $GW$ quasiparticle energies.
\\
\bigskip
\begin{center}
	\boxed{\includegraphics[width=0.5\linewidth]{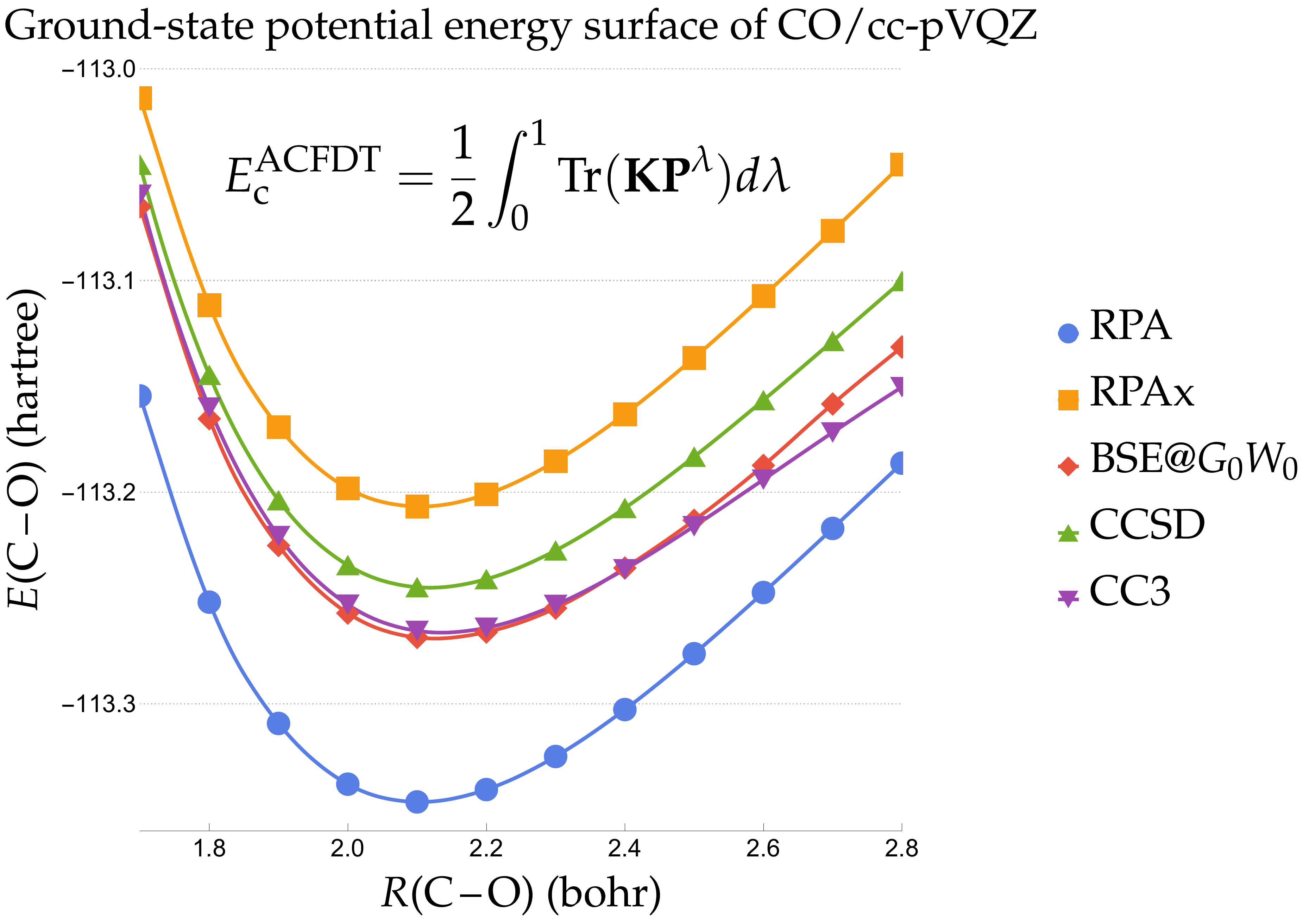}}
\end{center}
\bigskip
\end{abstract}

\maketitle


With a similar computational scaling as time-dependent density-functional theory (TD-DFT), \cite{Runge_1984,Casida} the many-body Green's function Bethe-Salpeter equation (BSE) formalism 
\cite{Salpeter_1951,Strinati_1988,Albrecht_1998,Rohlfing_1998,Benedict_1998,vanderHorst_1999} is a valuable alternative that has gained momentum in the past few years for studying molecular systems.\cite{Ma_2009,Pushchnig_2002,Tiago_2003,Palumno_2009,Rocca_2010,Sharifzadeh_2012,Cudazzo_2012,Boulanger_2014,Ljungberg_2015,Hirose_2015,Cocchi_2015,Ziaei_2017,Abramson_2017} 
It now stands as a cost-effective computational method that can model excited states \cite{Gonzales_2012,Loos_2020a} with a typical error of $0.1$--$0.3$ eV for spin-conserving transitions according to large and systematic benchmarks. \cite{Jacquemin_2015,Bruneval_2015,Hung_2016,Hung_2017,Krause_2017,Jacquemin_2017,Blase_2018}
One of the main advantages of BSE compared to TD-DFT is that it allows a faithful description of charge-transfer states. \cite{Lastra_2011,Blase_2011b,Baumeier_2012,Duchemin_2012,Cudazzo_2013,Ziaei_2016}
Moreover, when performed on top of a (partially) self-consistent {\evGW} calculation, \cite{Hybertsen_1986, Shishkin_2007, Blase_2011, Faber_2011,Rangel_2016,Kaplan_2016,Gui_2018} BSE@{\evGW} has been shown to be weakly dependent on its starting point (\eg, on the exchange-correlation functional selected for the underlying DFT calculation). \cite{Jacquemin_2015,Gui_2018}
However, similar to adiabatic TD-DFT, \cite{Levine_2006,Tozer_2000,Huix-Rotllant_2010,Elliott_2011} the static version of BSE cannot describe multiple excitations. \cite{Romaniello_2009a,Sangalli_2011,Loos_2019}

A significant limitation of the BSE formalism, as compared to TD-DFT, lies in the lack of analytical nuclear gradients (\ie, the first derivatives of the energy with respect to the nuclear displacements) for both the ground and excited states, \cite{Furche_2002} preventing efficient studies of excited-state processes (\eg, chemoluminescence and fluorescence) associated with geometric relaxation of ground and excited states, and structural changes upon electronic excitation. \cite{Bernardi_1996,Olivucci_2010,Navizet_2011,Robb_2007}
While calculations of the $GW$ quasiparticle energy ionic gradients is becoming increasingly popular,
\cite{Lazzeri_2008,Faber_2011b,Yin_2013,Faber_2015,Montserrat_2016,Zhenglu_2019} only one pioneering study of the excited-state BSE gradients has been published so far. \cite{Beigi_2003} 
In this seminal work devoted to small molecules (\ce{CO} and \ce{NH3}), only the BSE excitation energy gradients were calculated, while computing the Kohn-Sham (KS) LDA forces as its ground-state contribution.

In contrast to TD-DFT which relies on KS-DFT \cite{Hohenberg_1964,Kohn_1965,ParrBook} as its ground-state analog, the ground-state BSE energy is not a well-defined quantity, and no clear consensus has been found regarding its formal definition.
Consequently, the BSE ground-state formalism remains in its infancy with very few available studies for atomic and molecular systems. \cite{Olsen_2014,Holzer_2018,Li_2019,Li_2020}
In the largest available benchmark study \cite{Holzer_2018} encompassing the total energies of the atoms \ce{H}--\ce{Ne}, the atomization energies of the 26 small molecules forming the HEAT test set, \cite{Harding_2008} and the bond lengths and harmonic vibrational frequencies of $3d$ transition-metal monoxides, the BSE correlation energy, as evaluated within the adiabatic-connection fluctuation-dissipation theorem (ACFDT) framework, \cite{Furche_2005} was mostly discarded from the set of tested techniques due to instabilities (negative frequency modes in the BSE polarization propagator) and replaced by an approximate (RPAsX) approach where the screened-Coulomb potential matrix elements was removed from the resonant electron-hole contribution. \cite{Maggio_2016,Holzer_2018} 
Such a modified BSE polarization propagator was inspired by a previous study on the homogeneous electron gas (HEG). \cite{Maggio_2016} 
Within RPAsX, amounting to neglect excitonic effects in the electron-hole propagator, the question of using either KS-DFT or $GW$ eigenvalues in the construction of the propagator becomes further relevant, increasing accordingly the number of possible definitions for the ground-state correlation energy. 
Finally, renormalizing or not the Coulomb interaction by the interaction strength $\IS$ in the Dyson equation for the interacting polarizability (see below) leads to two different versions of the BSE correlation energy, \cite{Holzer_2018} emphasizing further the lack of general agreement around the definition of the ground-state BSE energy.

Here, in analogy to the random-phase approximation (RPA)-type formalisms \cite{Furche_2008,Toulouse_2009,Toulouse_2010,Angyan_2011,Ren_2012} and similarly to Refs.~\onlinecite{Olsen_2014,Maggio_2016,Holzer_2018}, the ground-state BSE energy is calculated in the adiabatic connection framework. 
Embracing this definition, the purpose of the present Letter is to investigate the quality of ground-state PES near equilibrium obtained within the BSE approach for several diatomic molecules.
The location of the minimum on the ground-state PES is of particular interest.
This study is a first necessary step towards the development of analytical nuclear gradients within the BSE@$GW$ formalism.
Thanks to comparisons with both similar and state-of-art computational approaches, we show that the ACFDT@BSE@$GW$ approach is surprisingly accurate, and can even compete with high-order coupled cluster (CC) methods in terms of absolute energies and equilibrium distances.
However, we also observe that, in some cases, unphysical irregularities on the ground-state PES, which are due to the appearance of a satellite resonance with a weight similar to that of the $GW$ quasiparticle peak. \cite{vanSetten_2015,Maggio_2017,Loos_2018,Veril_2018,Duchemin_2020}



In order to compute the neutral (optical) excitations of the system and their associated oscillator strengths, the BSE expresses the two-body propagator \cite{Strinati_1988, Martin_2016} 
\begin{multline}
\label{eq:BSE}
	\LBSE{}(1,2,1',2') = \LBSE{0}(1,2,1',2') 
	\\
	+ \int d3 d4 d5  d6  \LBSE{0}(1,4,1',3) \XiBSE{}(3,5,4,6) \LBSE{}(6,2,5,2')
\end{multline}
as the linear response of the one-body Green's function $\G{}$ with respect to a general non-local external potential
\begin{equation}
	\XiBSE{}(3,5,4,6) = i \fdv{[\vc{\Ha}(3) \delta(3,4) + \Sig{\xc}(3,4)]}{\G{}(6,5)},
\end{equation}
which takes into account the self-consistent variation of the Hartree potential 
\begin{equation}
	\vc{\Ha}(1) = - i \int d2 \vc{}(2) \G{}(2,2^+),
\end{equation}
(where $\vc{}$ is the bare Coulomb operator) and the exchange-correlation self-energy $\Sig{\xc}$.
In Eq.~\eqref{eq:BSE}, $\LBSE{0}(1,2,1',2') = - i \G{}(1,2')\G{}(2,1')$, and $(1)=(\br{1},\sigma_1,t_1)$ is a composite index gathering space, spin and time variables.
In the $GW$ approximation, \cite{Hedin_1965,Aryasetiawan_1998,Onida_2002,Martin_2016,Reining_2017} we have
\begin{equation}
	\SigGW{\xc}(1,2) = i \G{}(1,2) \W{}{}(1^+,2),
\end{equation}
where $\W{}{}$ is the screened Coulomb operator, and hence the BSE reduces to 
\begin{equation}
	\XiBSE{}(3,5,4,6) = \delta(3,4) \delta(5,6) \vc{}(3,6) - \delta(3,6) \delta(4,5) \W{}{}(3,4),
\end{equation}
where, as commonly done, we have neglected the term $\delta \W{}{}/\delta \G{}$ in the functional derivative of the self-energy. \cite{Hanke_1980,Strinati_1984,Strinati_1982}
Finally, the static approximation is enforced, \ie, $\W{}{}(1,2) = \W{}{}(\{\br{1}, \sigma_1, t_1\},\{\br{2}, \sigma_2, t_2\}) \delta(t_1-t_2)$, which corresponds to restricting $\W{}{}$ to its static limit, \ie, $\W{}{}(1,2) = \W{}{}(\{\br{1},\sigma_1\},\{\br{2},\sigma_2\}; \omega=0)$.


For a closed-shell system in a finite basis, to compute the singlet BSE excitation energies (within the static approximation) of the physical system (\ie, $\IS = 1$), one must solve the following linear response problem \cite{Casida,Dreuw_2005,Martin_2016} 
\begin{equation}
\label{eq:LR}
	\begin{pmatrix}
		\bA{\IS}	&	\bB{\IS}	\\
		-\bB{\IS}	&	-\bA{\IS}	\\
	\end{pmatrix}
	\begin{pmatrix}
		\bX{\IS}_m	\\
		\bY{\IS}_m	\\
	\end{pmatrix}
	=
	\Om{m}{\IS}
	\begin{pmatrix}
		\bX{\IS}_m	\\
		\bY{\IS}_m	\\
	\end{pmatrix},
\end{equation}
where $\Om{m}{\IS}$ is the $m$th excitation energy with eigenvector $\T{(\bX{\IS}_m \, \bY{\IS}_m)}$ at interaction strength $\IS$, $\T{}$ is the matrix transpose, and we assume real-valued spatial orbitals $\{\MO{p}(\br{})\}_{1 \le p \le \Norb}$.
The matrices $\bA{\IS}$, $\bB{\IS}$, $\bX{\IS}$, and $\bY{\IS}$ are all of size $\Nocc \Nvir \times \Nocc \Nvir$ where $\Nocc$ and $\Nvir$ are the number of occupied and virtual orbitals (\ie, $\Norb = \Nocc + \Nvir$ is the total number of spatial orbitals), respectively.
In the following, the index $m$ labels the $\Nocc \Nvir$ single excitations, $i$ and $j$ are occupied orbitals, $a$ and $b$ are unoccupied orbitals, while $p$, $q$, $r$, and $s$ indicate arbitrary orbitals.

In the absence of instabilities (\ie, when $\bA{\IS} - \bB{\IS}$ is positive-definite), \cite{Dreuw_2005} Eq.~\eqref{eq:LR} is usually transformed into an Hermitian eigenvalue problem of smaller dimension
\begin{equation}
\label{eq:small-LR}
	(\bA{\IS} - \bB{\IS})^{1/2} (\bA{\IS} + \bB{\IS}) (\bA{\IS} - \bB{\IS})^{1/2} \bZ{m}{\IS} = (\Om{m}{\IS})^2 \bZ{m}{\IS},
\end{equation}
where the excitation amplitudes are
\begin{subequations}
\begin{align}
	(\bX{\IS} + \bY{\IS})_m = (\Om{m}{\IS})^{-1/2} (\bA{\IS} - \bB{\IS})^{+1/2} \bZ{m}{\IS},
	\\
	(\bX{\IS} - \bY{\IS})_m = (\Om{m}{\IS})^{+1/2} (\bA{\IS} - \bB{\IS})^{-1/2} \bZ{m}{\IS}.
\end{align}
\end{subequations}
Introducing the so-called Mulliken notation for the bare two-electron integrals
\begin{equation}
	\ERI{pq}{rs} = \iint \frac{\MO{p}(\br{}) \MO{q}(\br{}) \MO{r}(\br{}') \MO{s}(\br{}')}{\abs*{\br{} - \br{}'}} \dbr{} \dbr{}',
\end{equation}
and the corresponding (static) screened Coulomb potential matrix elements at coupling strength $\IS$ 
\begin{equation}
	\W{pq,rs}{\IS} = \iint  \MO{p}(\br{}) \MO{q}(\br{}) \W{}{\IS}(\br{},\br{}') \MO{r}(\br{}') \MO{s}(\br{}')  \dbr{} \dbr{}',
\end{equation}
the BSE matrix elements read
\begin{subequations}
\begin{align}
	\label{eq:LR_BSE-A}
	\ABSE{ia,jb}{\IS} & = \delta_{ij} \delta_{ab} (\eGW{a} - \eGW{i}) + \IS \qty[ 2 \ERI{ia}{jb}   -  \W{ij,ab}{\IS} ],
	\\ 
	\label{eq:LR_BSE-B}
	\BBSE{ia,jb}{\IS} & =  \IS \qty[ 2 \ERI{ia}{bj} -   \W{ib,aj}{\IS} ],
\end{align}
\end{subequations}
where $\eGW{p}$ are the $GW$ quasiparticle energies. 
In the standard BSE approach, $\W{}{\IS}$ is built within the direct RPA scheme, \ie,
\begin{subequations}
\label{eq:wrpa}
\begin{align}
	\W{}{\IS}(\br{},\br{}') 
	& = \int \frac{\epsilon_{\IS}^{-1}(\br{},\br{}''; \omega=0)}{\abs*{\br{}' - \br{}''}} \dbr{}'' , 
	\\ 
	\epsilon_{\IS}(\br{},\br{}'; \omega) 
	& = \delta(\br{}-\br{}') - \IS  \int \frac{\chi_{0}(\br{},\br{}''; \omega)}{\abs*{\br{}' - \br{}''}} \dbr{}'' ,
\end{align}
\end{subequations}
with $\epsilon_{\IS}$ the dielectric function at coupling constant $\IS$ and $\chi_{0}$ the non-interacting polarizability.  In the occupied-to-virtual orbital product basis, the spectral representation of $\W{}{\IS}$ can be written as follows in the case of real spatial orbitals
\begin{multline}
\label{eq:W}
	\W{ij,ab}{\IS}(\omega) = \ERI{ij}{ab} +  2 \sum_m^{\Nocc \Nvir} \sERI{ij}{m}{\IS} \sERI{ab}{m}{\IS}
	\\
	\times  \qty(\frac{1}{\omega - \OmRPA{m}{\IS} + i \eta} - \frac{1}{\omega + \OmRPA{m}{\IS} - i \eta}),
\end{multline}
where the spectral weights at coupling strength $\IS$ read
\begin{equation}
	\sERI{pq}{m}{\IS} =  \sum_i^{\Nocc} \sum_a^{\Nvir} \ERI{pq}{ia} (\bX{\IS}_m + \bY{\IS}_m)_{ia}.
\end{equation}
In the case of complex orbitals, we refer the reader to Ref.~\onlinecite{Holzer_2019} for a correct use of complex conjugation in the spectral representation of $\W{}{}$.
\titou{Note that, in the case of {\GOWO}, the RPA neutral excitations in Eq.~\eqref{eq:W} are computed using the HF orbital energies.}

In Eq.~\eqref{eq:W}, $\eta$ is a positive infinitesimal, and $\OmRPA{m}{\IS}$ are the direct (\ie, without exchange) RPA neutral excitation energies computed by solving the linear eigenvalue problem \eqref{eq:LR} with the following matrix elements
\begin{subequations}
\begin{align}
	\label{eq:LR_RPA-A}
	\ARPA{ia,jb}{\IS} & = \delta_{ij} \delta_{ab} (\eHF{a} - \eHF{i}) +   2 \IS \ERI{ia}{jb},
	\\ 
	\label{eq:LR_RPA-B}
	\BRPA{ia,jb}{\IS} & =   2 \IS \ERI{ia}{bj},
\end{align}
\end{subequations}
where $\eHF{p}$ are the Hartree-Fock (HF) orbital energies.

The relationship between the BSE formalism and the well-known RPAx (\ie, RPA with exchange) approach can be obtained by switching off the screening so that $\W{}{\IS}$ reduces to the bare Coulomb potential $\vc{}$. 
In this limit, the $GW$ quasiparticle energies reduce to the HF eigenvalues, and Eqs.~\eqref{eq:LR_BSE-A} and \eqref{eq:LR_BSE-B} to the RPAx equations: 
\begin{subequations}
\begin{align}
	\label{eq:LR_RPAx-A}
	\ARPAx{ia,jb}{\IS} & = \delta_{ij} \delta_{ab} (\eHF{a} - \eHF{i}) + \IS \qty[ 2 \ERI{ia}{jb} -    \ERI{ij}{ab} ],
	\\ 
	\label{eq:LR_RPAx-B}
	\BRPAx{ia,jb}{\IS} & =  \IS \qty[ 2 \ERI{ia}{bj} - \ERI{ib}{aj} ].
\end{align}
\end{subequations}

The key quantity to define in the present context is the total BSE ground-state energy $\EBSE$.
Although this choice is not unique, \cite{Holzer_2018} we propose here to define it as
\begin{equation}
\label{eq:EtotBSE}
	\EBSE = \Enuc + \EHF + \EcBSE,
\end{equation}
where $\Enuc$ and $\EHF$ are the nuclear repulsion energy and electronic ground-state HF energy (respectively), and
\begin{equation}
\label{eq:EcBSE}
	\EcBSE = \frac{1}{2} \int_0^1 \Tr(\bK \bP{\IS}) d\IS
\end{equation}
is the ground-state BSE correlation energy computed in the adiabatic connection framework, where 
\begin{equation}
\label{eq:K}
	\bK = 
	\begin{pmatrix}
		\btA{\IS=1}	&	\bB{\IS=1}	\\
		\bB{\IS=1}	&	\btA{\IS=1}	\\
	\end{pmatrix}
\end{equation}
is the interaction kernel \cite{Angyan_2011, Holzer_2018} [with $\tA{ia,jb}{\IS} = \titou{2} \IS \ERI{ia}{bj}$],
\begin{equation}
\label{eq:2DM}
	\bP{\IS} = 
	\begin{pmatrix}
		\bY{\IS} \T{(\bY{\IS})}		&	\bY{\IS} \T{(\bX{\IS})}	\\
		\bX{\IS} \T{(\bY{\IS})}		&	\bX{\IS} \T{(\bX{\IS})}	\\
	\end{pmatrix}
	-
	\begin{pmatrix}
		\bO		&	\bO	\\
		\bO		&	\bI	\\
	\end{pmatrix}
\end{equation} 
is the correlation part of the two-electron density matrix at interaction strength $\IS$, and $\Tr$ denotes the matrix trace.
Note that the present definition of the BSE correlation energy [see Eq.~\eqref{eq:EcBSE}], which we refer to as BSE@$GW$@HF in the following, has been named ``XBS'' for ``extended Bethe Salpeter'' by Holzer \textit{et al.} \cite{Holzer_2018}
\titou{For the sake of completeness, comparisons between the extended and regular BSE schemes can be found in the {\SI}.}
In contrast to DFT where the electron density is fixed along the adiabatic path, in the present formalism, the density is not maintained as $\IS$ varies.
Therefore, an additional contribution to Eq.~\eqref{eq:EcBSE} originating from the variation of the Green's function along the adiabatic connection should be, in principle, added.
However, as commonly done within RPA and RPAx, \cite{Toulouse_2009, Toulouse_2010, Colonna_2014, Holzer_2018} we shall neglect it in the present study. 

Equation \eqref{eq:EcBSE} can also be straightforwardly applied to RPA and RPAx, the only difference being the expressions of $\bA{\IS}$ and $\bB{\IS}$ used to obtain the eigenvectors $\bX{\IS}$ and $\bY{\IS}$ entering in the definition of $\bP{\IS}$ [see Eq.~\eqref{eq:2DM}].
For RPA, these expressions have been provided in Eqs.~\eqref{eq:LR_RPA-A} and \eqref{eq:LR_RPA-B}, and their RPAx analogs in Eqs.~\eqref{eq:LR_RPAx-A} and \eqref{eq:LR_RPAx-B}.
In the following, we will refer to these two types of calculations as RPA@HF and RPAx@HF, respectively.
Finally, we will also consider the RPA@$GW$@HF scheme which consists in replacing the HF orbital energies in Eq.~\eqref{eq:LR_RPA-A} by the $GW$ quasiparticles energies.

Note that, for spin-restricted closed-shell molecular systems around their equilibrium geometry (such as the ones studied here), one rarely encounters singlet instabilities as these systems can be classified as weakly correlated.
However, singlet instabilities may appear in the presence of strong correlation, \eg, when the bonds are stretched, hampering in particular the calculation of atomization energies. \cite{Holzer_2018}
Even for weakly correlated systems, triplet instabilities are much more common, but triplet excitations do not contribute to the correlation energy in the ACFDT formulation. \cite{Toulouse_2009, Toulouse_2010, Angyan_2011}

\titou{The restricted HF formalism has been systematically employed in the present study.}
All the $GW$ calculations performed to obtain the screened Coulomb operator and the quasiparticle energies are done using a (restricted) HF starting point, which is an adequate choice in the case of the (small) systems that we have considered here.
Perturbative $GW$ (or {\GOWO}) \cite{Hybertsen_1985a, Hybertsen_1986} calculations are employed as starting points to compute the BSE neutral excitations.
In the case of {\GOWO}, the quasiparticle energies are obtained by linearizing the frequency-dependent quasiparticle equation.
Further details about our implementation of {\GOWO} can be found in Refs.~\onlinecite{Loos_2018, Veril_2018}.
Finally, the infinitesimal $\eta$ is set to zero for all calculations.
The numerical integration required to compute the correlation energy along the adiabatic path [see Eq.~\eqref{eq:EcBSE}] is performed with a 21-point Gauss-Legendre quadrature. 
Comparison with the so-called plasmon (or trace) formula \cite{Furche_2008} at the RPA level has confirmed the excellent accuracy of this quadrature scheme over $\IS$. 

For comparison purposes, we have also computed the PES at the second-order M{\o}ller-Plesset perturbation theory (MP2), as well as with various increasingly accurate CC methods, namely, CC2 \cite{Christiansen_1995}, CCSD, \cite{Purvis_1982} and CC3. \cite{Christiansen_1995b} 
These calculations have been performed with DALTON \cite{dalton} and PSI4. \cite{Psi4}
The computational cost of these methods, in their usual implementation, scale as $\order*{N^5}$, $\order*{N^5}$, $\order*{N^6}$, and $\order*{N^7}$, respectively.
As shown in Refs.~\onlinecite{Hattig_2005c,Budzak_2017}, CC3 provides extremely accurate ground-state (and excited-state) geometries, and will be taken as reference in the present study.
\titou{In order to check further the overall accuracy of CC3, we have performed CCSDT and CCSDT(Q) calculations \cite{MRCC} at equilibrium bond lengths.
These results are provided in the {\SI} and clearly evidenced the excellent accuracy of CC3, the maximum absolute deviation between CC3 and CCSDT(Q) being $0.2\%$ at equilibrium.}
All the other calculations have been performed with our locally developed $GW$ software. \cite{Loos_2018,Veril_2018}
As one-electron basis sets, we employ the Dunning family (cc-pVXZ) defined with cartesian Gaussian functions. 
Unless otherwise stated, the frozen-core approximation is not applied in order to provide a fair comparison between methods.
We have, however, found that our conclusions hold within the frozen-core approximation (see the {\SI} for 	 information).

Because Eq.~\eqref{eq:EcBSE} requires the entire BSE singlet excitation spectrum for each quadrature point, we perform several complete diagonalization of the $\Nocc \Nvir \times \Nocc \Nvir$ BSE linear response matrix [see Eq.~\eqref{eq:small-LR}], which corresponds to a $\order{\Nocc^3 \Nvir^3} = \order{\Norb^6}$ computational cost.
This step is, by far, the computational bottleneck in our current implementation.
However, we are currently pursuing different avenues to lower the formal scaling and practical cost of this step by computing the two-electron density matrix of Eq.~\eqref{eq:2DM} via a quadrature in frequency space. \cite{Duchemin_2019,Duchemin_2020}

\begin{figure*}
	\includegraphics[width=0.49\linewidth]{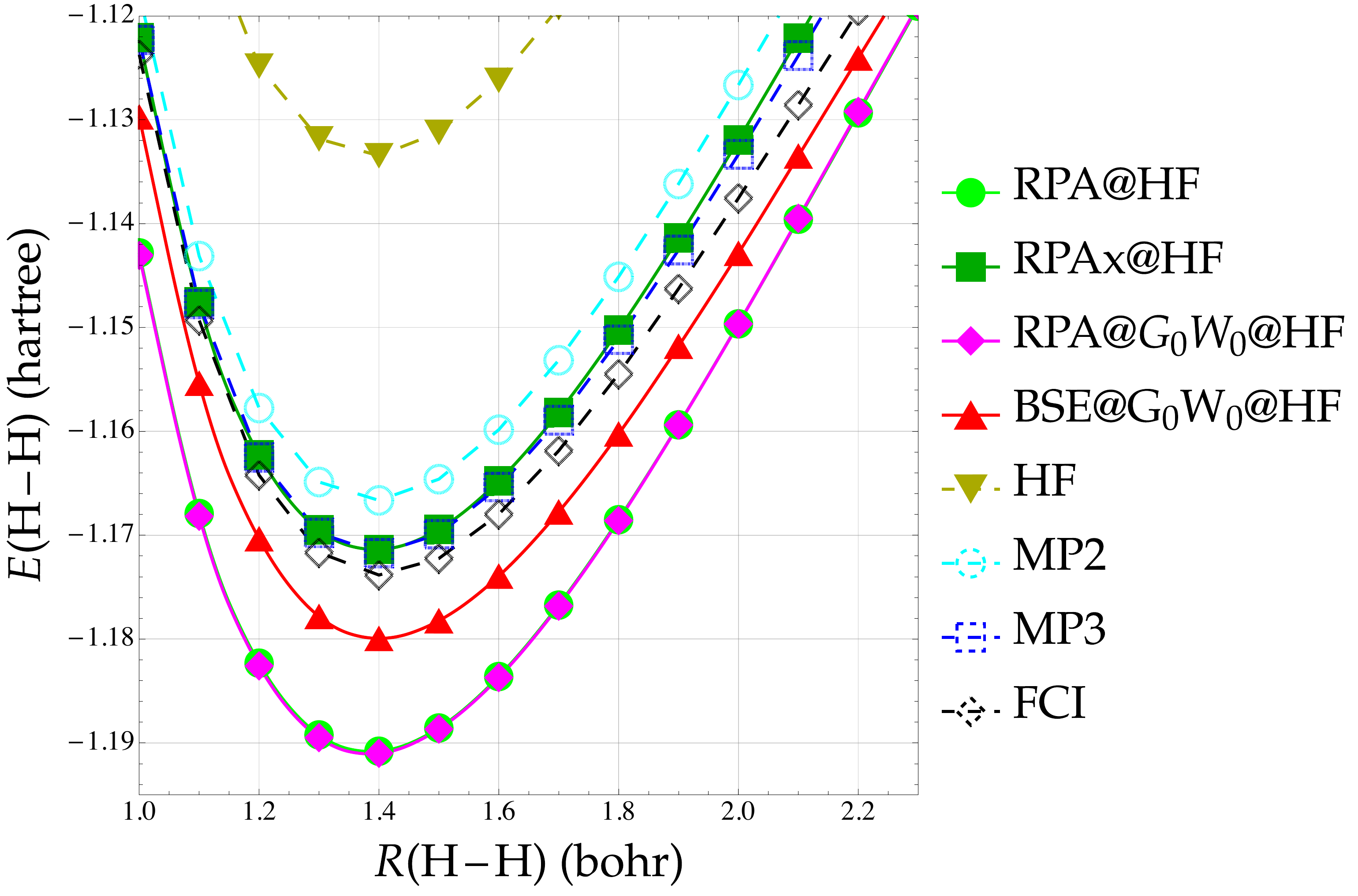}
	\includegraphics[width=0.49\linewidth]{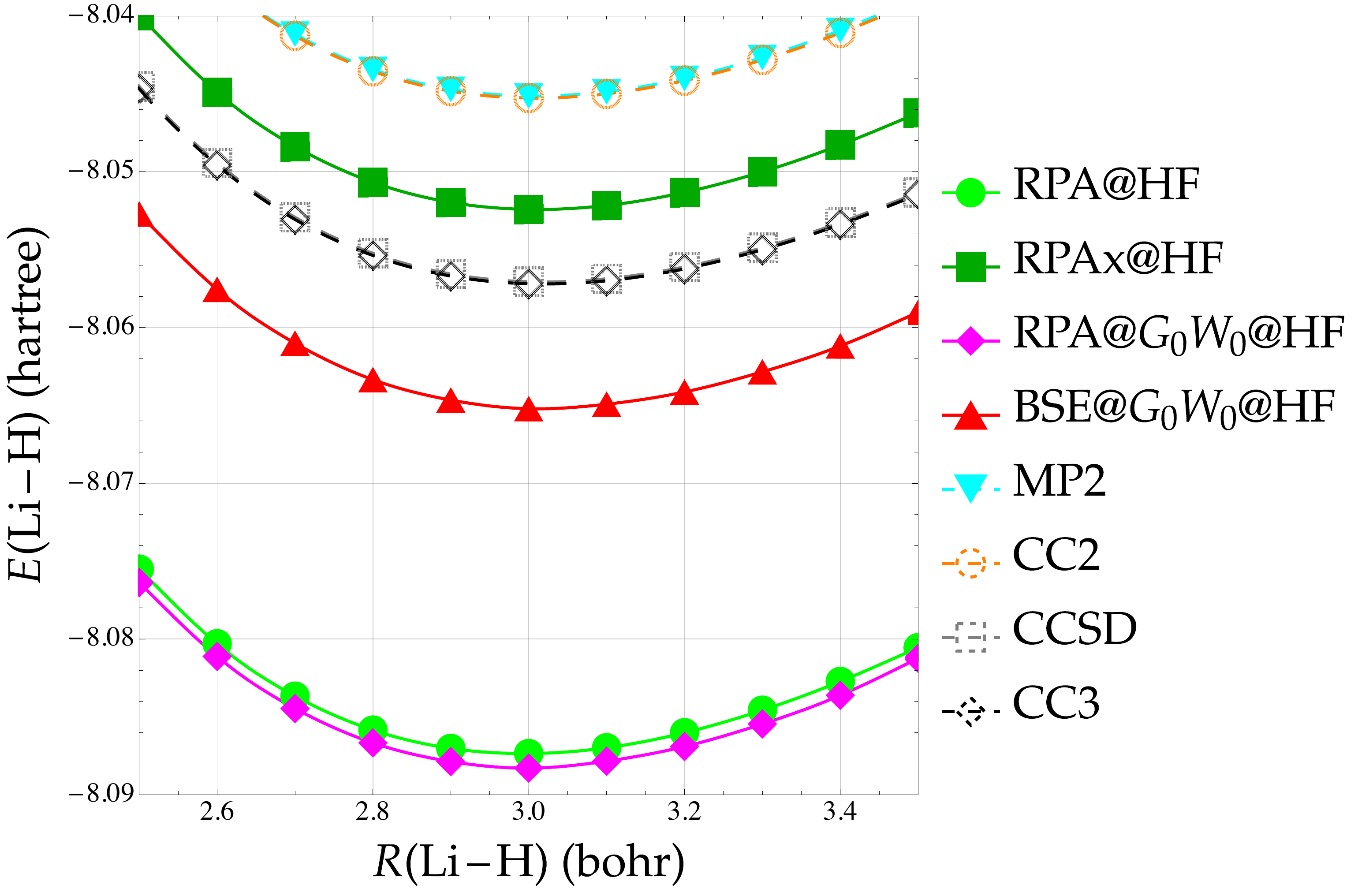}
\caption{
Ground-state PES of \ce{H2} (left) and \ce{LiH} (right) around their respective equilibrium geometry obtained at various levels of theory with the cc-pVQZ basis set. 
\label{fig:PES-H2-LiH}
}
\end{figure*}

\begin{figure*}
	\includegraphics[height=0.35\linewidth]{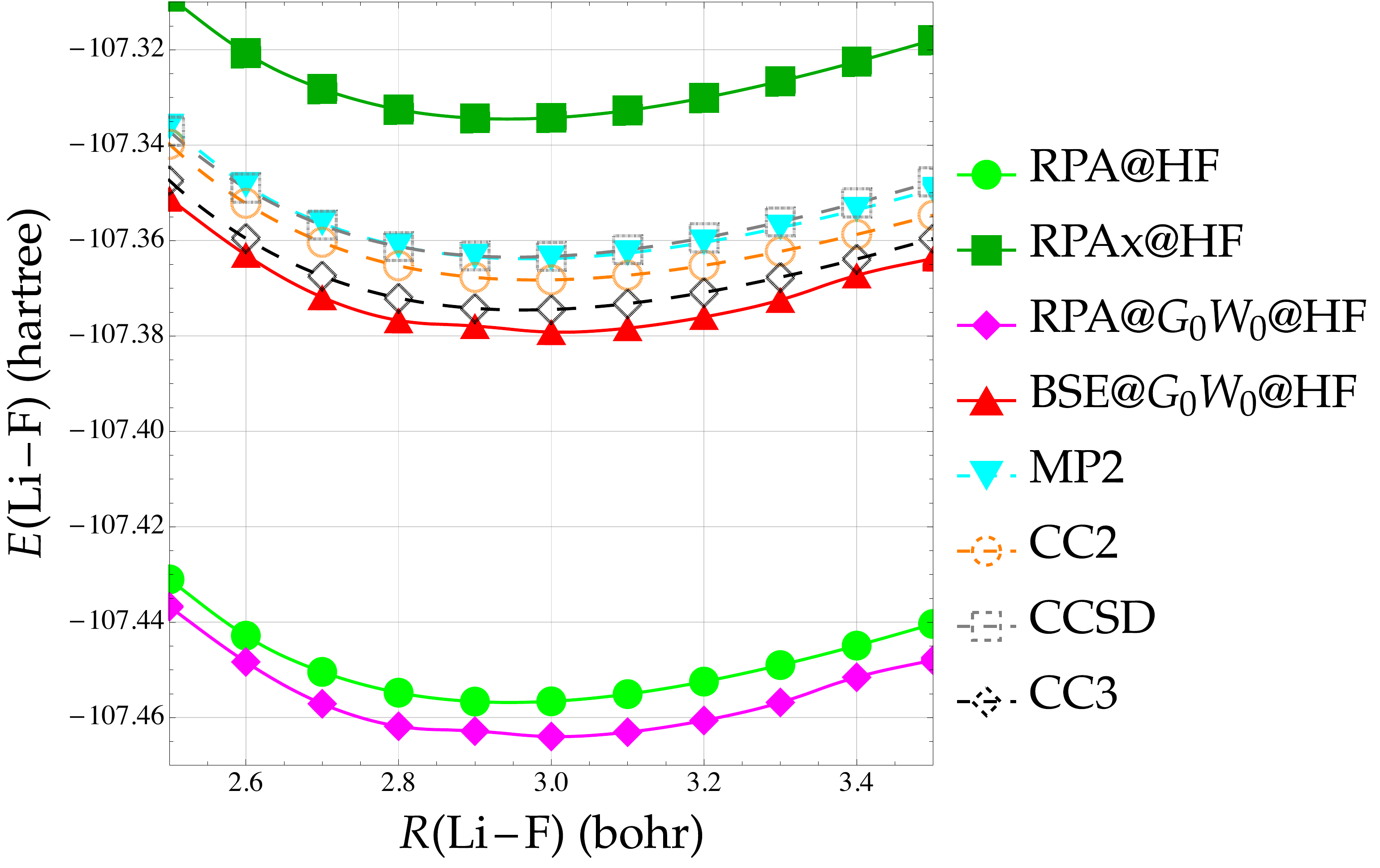}
	\includegraphics[height=0.35\linewidth]{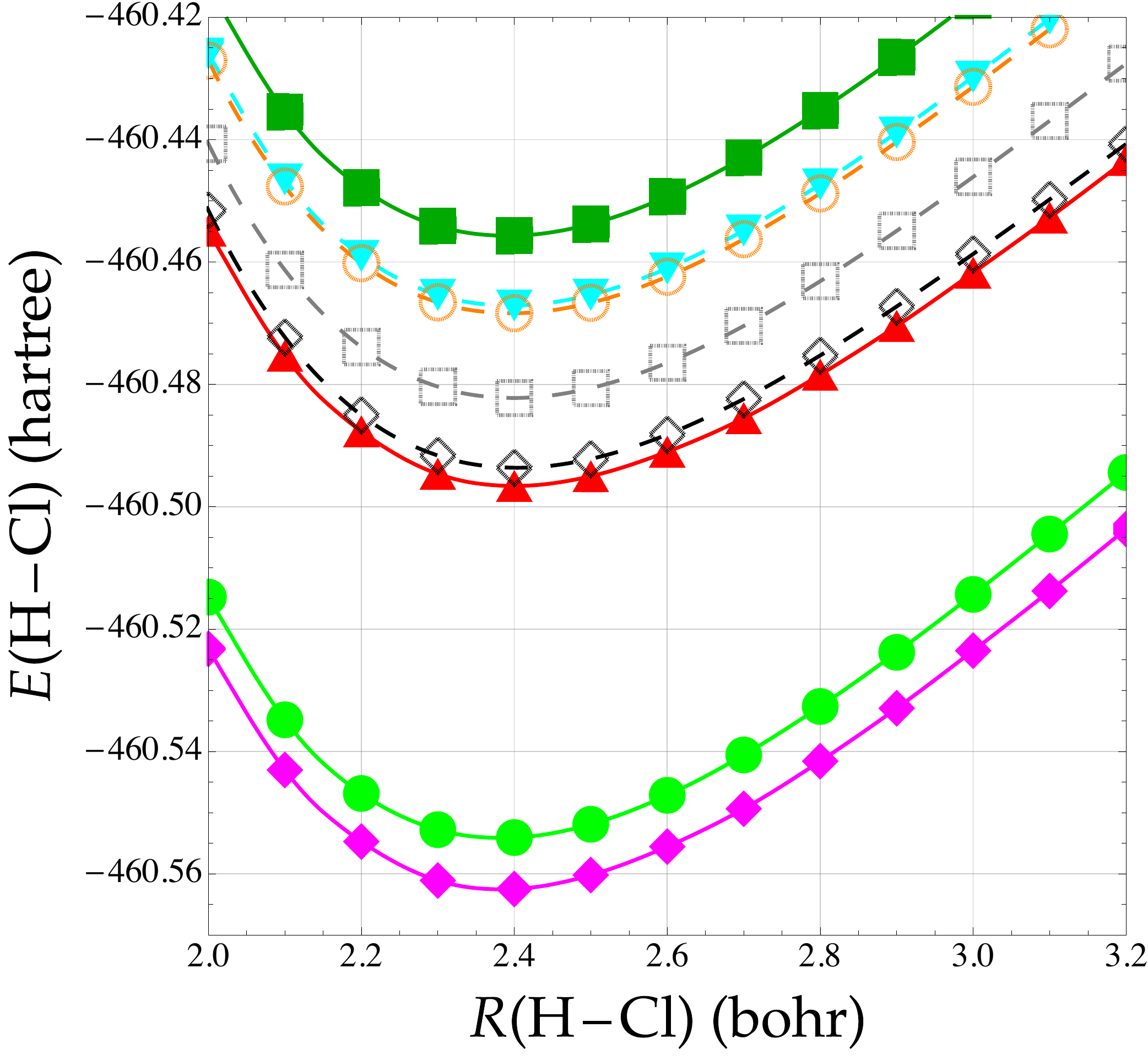}
\caption{
Ground-state PES of \ce{LiF} (left) and \ce{HCl} (right) around their respective equilibrium geometry obtained at various levels of theory with the cc-pVQZ basis set. 
\label{fig:PES-LiF-HCl}
}
\end{figure*}

\begin{figure*}
	\includegraphics[height=0.26\linewidth]{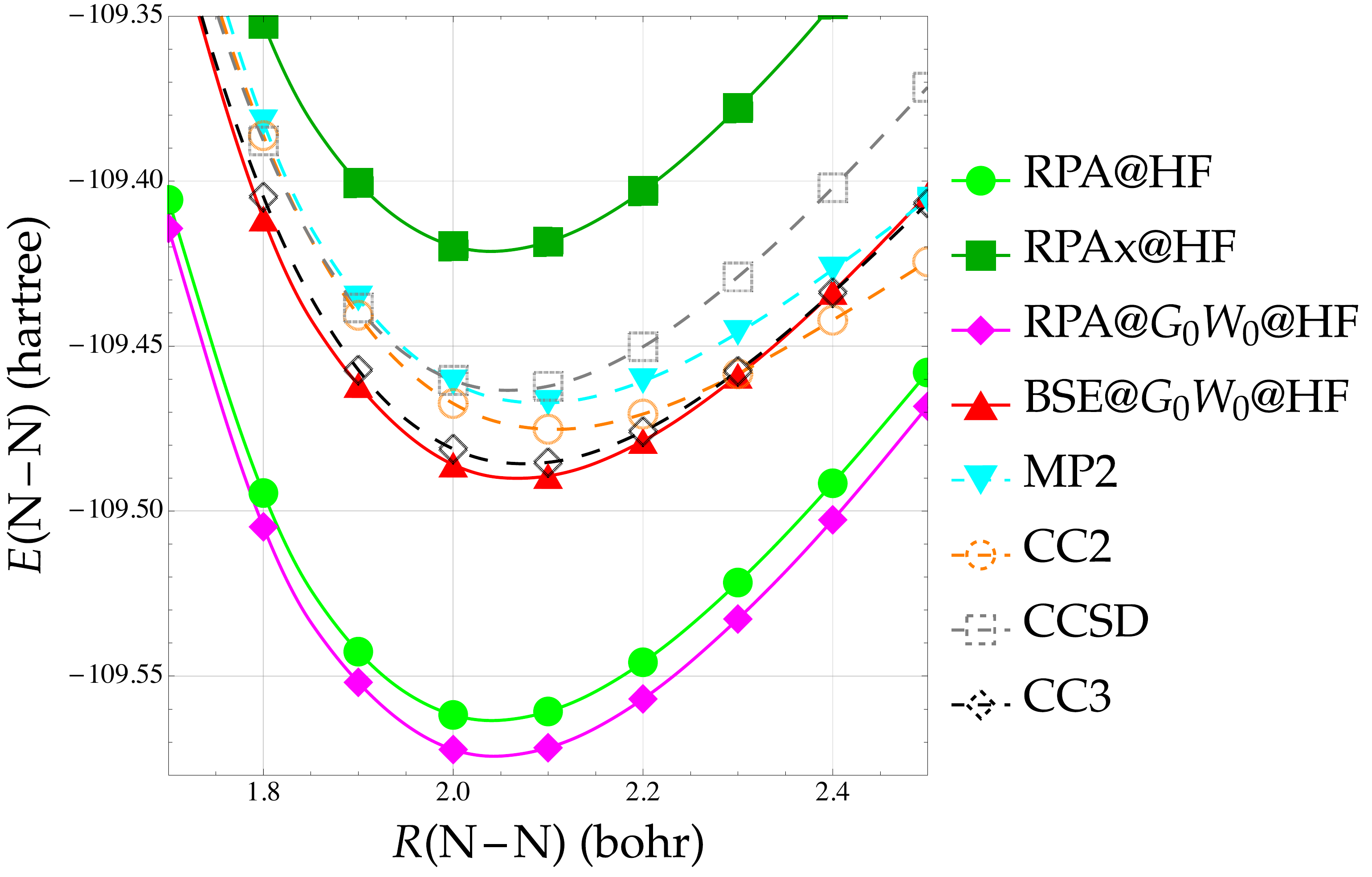}
	\includegraphics[height=0.26\linewidth]{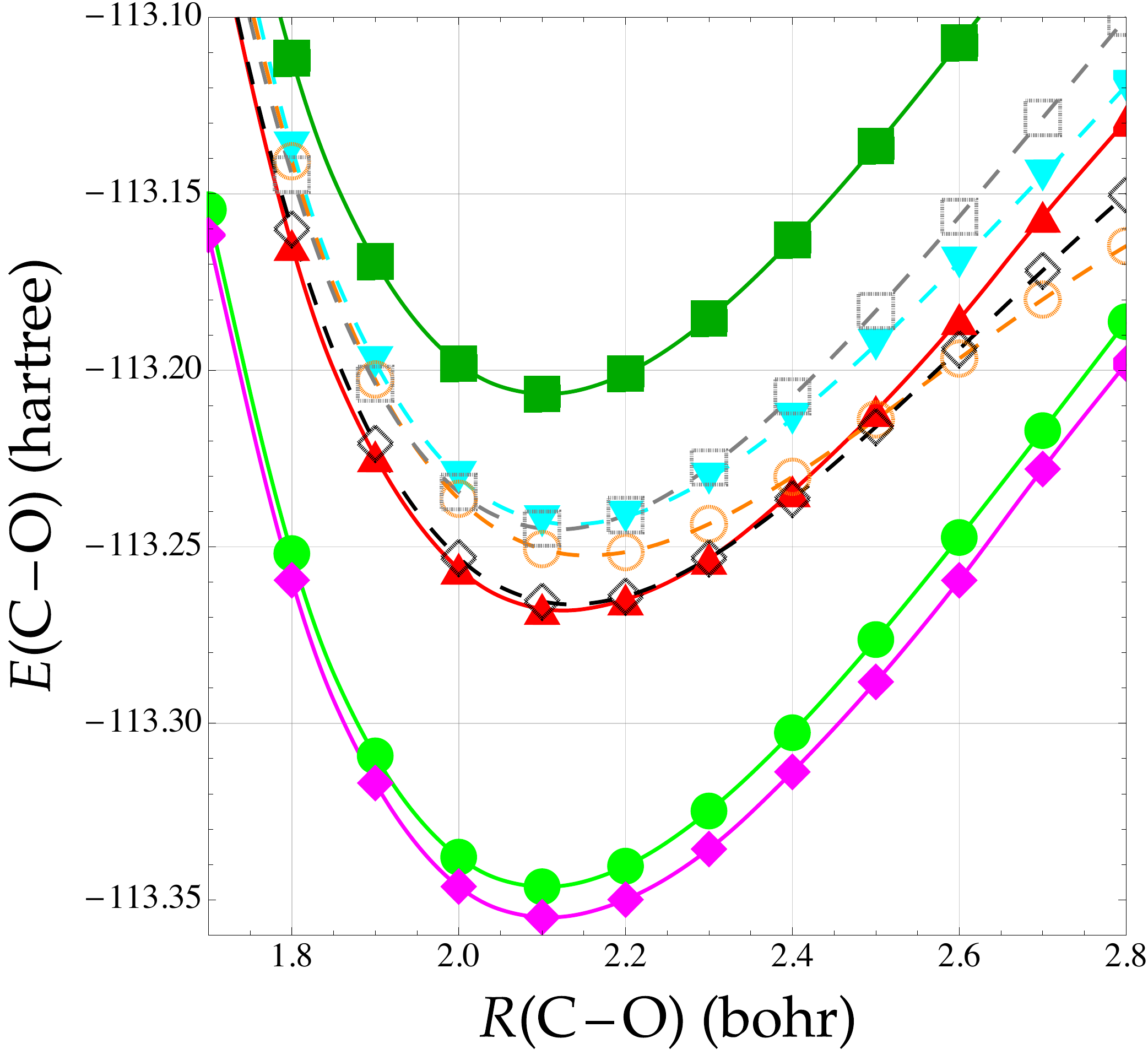}
	\includegraphics[height=0.26\linewidth]{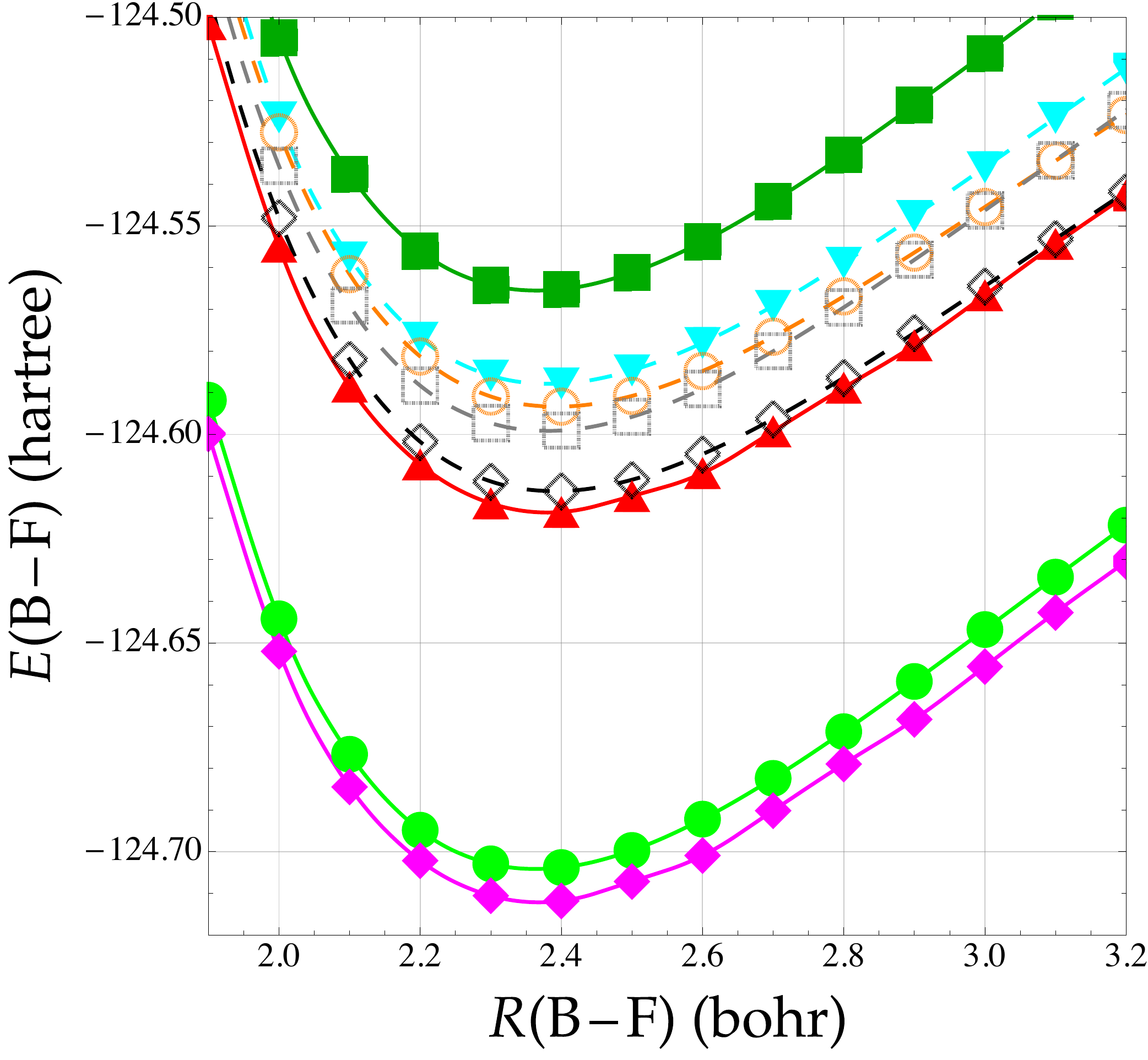}
\caption{
Ground-state PES of the isoelectronic series \ce{N2} (left), \ce{CO} (center), and \ce{BF} (right) around their respective equilibrium geometry obtained at various levels of theory with the cc-pVQZ basis set. 
\label{fig:PES-N2-CO-BF}
}
\end{figure*}

\begin{figure}
	\includegraphics[width=\linewidth]{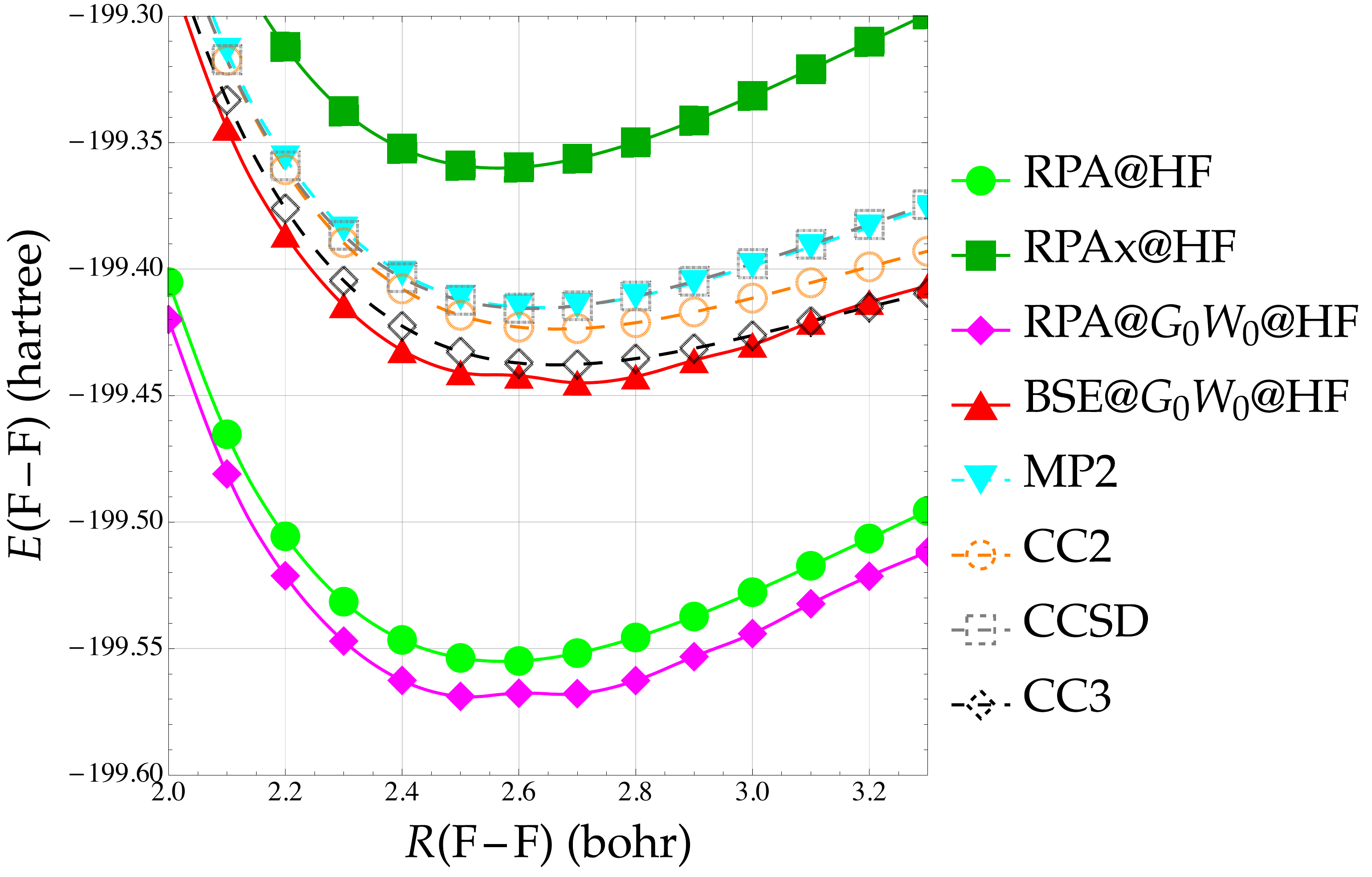}
\caption{
Ground-state PES of \ce{F2} around its equilibrium geometry obtained at various levels of theory with the cc-pVQZ basis set. 
\label{fig:PES-F2}
}
\end{figure}

In order to illustrate the performance of the BSE-based adiabatic connection formulation, we compute the ground-state PES of several closed-shell diatomic molecules around their equilibrium geometry: \ce{H2}, \ce{LiH}, \ce{LiF}, \ce{HCl}, \ce{N2}, \ce{CO}, \ce{BF}, and \ce{F2}.
The PES of these molecules are represented in Figs.~\ref{fig:PES-H2-LiH}, \ref{fig:PES-LiF-HCl}, \ref{fig:PES-N2-CO-BF}, and \ref{fig:PES-F2}, while the computed equilibrium distances and correlation energies are gathered in Table \ref{tab:Req}.
Both of these properties are computed with Dunning's cc-pVQZ basis set. 
Graphs and tables for the corresponding double- and triple-$\zeta$ basis sets can be found in the {\SI}.

\begin{squeezetable}
\begin{table*}
\caption{
Equilibrium bond length $\Req$ (in bohr) and correlation energy $\Ec$ (in millihartree) for the ground state of diatomic molecules obtained with the cc-pVQZ basis set at various levels of theory. 
For each system and each method, the correlation energy is computed at its respective equilibrium bond length (\ie, $R = \Req$).
When irregularities appear in the PES, the $\Req$ values are reported in parenthesis and they have been obtained by fitting a Morse potential to the PES.
The error (in \%) compared to the reference CC3 values are reported in square brackets.
}
\label{tab:Req}
	\begin{ruledtabular}
	\begin{tabular}{lllllllll}
					&	\mc{8}{c}{Equilibrium bond length $\Req$ (bohr) }	\\
									\cline{2-9}
	Method			&	\tabc{\ce{H2}\fnm[1]}		&	\tabc{\ce{LiH}}		&	\tabc{\ce{LiF}}		&	\tabc{\ce{HCl}}		&	\tabc{\ce{N2}}		&	\tabc{\ce{CO}}		&	\tabc{\ce{BF}}		&	\tabc{\ce{F2}}				\\
	\hline
	CC3				&	1.402				&	3.019				&	2.963				&	2.403				&	2.075				&	2.136				&	2.390				&	2.663				\\
	CCSD			&	1.402[$+0.0\%$]	&	3.020[$+0.0\%$]	&	2.953[$-0.3\%$]		&	2.398[$-0.2\%$]	&	2.059[$-0.8\%$]	&	2.118[$-0.8\%$]	&	2.380[$-0.4\%$]	&	2.621[$-1.6\%$]	\\
	CC2				&	1.391[$-0.8\%$]	&	3.010[$-0.3\%$]	&	2.982[$+0.6\%$]		&	2.396[$-0.3\%$]	&	2.106[$+1.5\%$]	&	2.156[$+0.9\%$]	&	2.393[$+0.1\%$]	&	2.665[$+0.1\%$]	\\
	MP2				&	1.391[$-0.8\%$]	&	3.008[$-0.4\%$]	&	2.970[$+0.2\%$]		&	2.395[$-0.3\%$]	&	2.091[$+0.8\%$]	&	2.137[$+0.1\%$]	&	2.382[$-0.3\%$]	&	2.634[$-1.1\%$]	\\
	BSE@{\GOWO}@HF	&	1.399[$-0.2\%$]	&	3.017[$-0.1\%$]	&	(2.973)[$+0.3\%$]	&	2.400[$-0.1\%$]	&	2.065[$-0.5\%$]	&	2.134[$-0.1\%$]	&	2.385[$-0.2\%$]	&	(2.638)[$-0.9\%$]	\\
	RPA@{\GOWO}@HF	&	1.382[$-1.4\%$]	&	2.997[$-0.7\%$]	&	(2.965)[$+0.1\%$]	&	2.370[$-1.5\%$]	&	2.043[$-1.5\%$]	&	2.132[$-0.2\%$]	&	2.365[$-1.1\%$]	&	(2.571)[$-3.5\%$]	\\
	RPAx@HF			&	1.394[$-0.6\%$]	&	3.011[$-0.3\%$]	&	2.944[$-0.6\%$]		&	2.391[$-0.5\%$]	&	2.041[$-1.6\%$]	&	2.104[$-1.5\%$]	&	2.366[$-1.0\%$]	&	2.565[$-3.7\%$]	\\
	RPA@HF			&	1.386[$-1.1\%$]	&	2.994[$-0.8\%$]	&	2.946[$-0.6\%$]		&	2.382[$-0.9\%$]	&	2.042[$-1.6\%$]	&	2.103[$-1.5\%$]	&	2.364[$-1.1\%$]	&	2.573[$-3.4\%$]	\\
\hline
					&	\mc{8}{c}{Correlation energy $-\Ec$ (millihartree)}	\\
									\cline{2-9}
	Method			&	\tabc{\ce{H2}\fnm[1]}		&	\tabc{\ce{LiH}}		&	\tabc{\ce{LiF}}		&	\tabc{\ce{HCl}}		&	\tabc{\ce{N2}}		&	\tabc{\ce{CO}}		&	\tabc{\ce{BF}}		&	\tabc{\ce{F2}}				\\
	\hline
	CC3				&	40.4			&	70.0			&	383.7				&	382.2			&	494.4			&	477.6			&	447.5				&	668.9				\\
	CCSD			&	40.4[$+0.0\%$]	&	69.8[$-0.2\%$]	&	372.6[$-2.9\%$]		&	370.8[$-3.0\%$]	&	470.6[$-4.8\%$]	&	455.2[$-4.7\%$]	&	432.9[$-3.3\%$]	&	644.0[$-3.7\%$]	\\
	CC2				&	33.3[$-17.6\%$]	&	57.2[$-18.1\%$]	&	376.7[$-1.8\%$]		&	356.9[$-6.6\%$]	&	488.0[$-1.3\%$]	&	465.5[$-2.5\%$]	&	427.3[$-4.5\%$]	&	654.9[$-2.1\%$]	\\
	MP2				&	33.2[$-17.9\%$]	&	57.9[$-17.2\%$]	&	373.0[$-2.8\%$]		&	355.7[$-6.9\%$]	&	478.0[$-3.3\%$]	&	455.0[$-4.7\%$]	&	421.6[$-5.8\%$]	&	644.3[$-3.7\%$]	\\
	BSE@{\GOWO}@HF	&	46.5[$+15.1\%$]	&	78.0[$+11.4\%$]	&	388.3[$+1.2\%$]		&	385.1[$+0.8\%$]	&	497.9[$0.7+\%$]	&	480.0[$+0.5\%$]	&	452.3[$+1.1\%$]	&	673.9[$0.8+\%$]	\\
	RPA@{\GOWO}@HF	&	57.6[$+42.6\%$]	&	101.1[$+44.5\%$]&	473.1[$+23.3\%$]	&	451.2[$+18.1\%$]&	580.3[$+17.4\%$]&	566.5[$+18.6\%$]&	545.5[$+21.9\%$]	&	794.3[$+18.8\%$]	\\
	RPAx@HF			&	37.9[$-6.2\%$]	&	65.2[$-6.8\%$]	&	343.6[$-10.5\%$]	&	344.2[$-9.9\%$]	&	427.2[$-13.6\%$]&	416.3[$-12.8\%$]&	399.1[$-10.8\%$]	&	586.1[$-12.4\%$]	\\
	RPA@HF			&	57.3[$+42.0\%$]	&	100.2[$+43.2\%$]&	465.9[$+21.4\%$]	&	442.7[$+15.8\%$]&	569.4[$+15.2\%$]&	555.9[$+16.4\%$]&	537.7[$+20.2\%$]	&	781.3[$+16.8\%$]	\\
	\end{tabular}
	\end{ruledtabular}
\fnt[1]{\titou{For \ce{H2}, both CC3 and CCSD are equivalent to FCI.}}
\end{table*}
\end{squeezetable}

Let us start with the two smallest molecules, \ce{H2} and \ce{LiH}.
Their PES computed with the cc-pVQZ basis are reported in Fig.~\ref{fig:PES-H2-LiH}.
For \ce{H2}, we take as reference the full configuration interaction (FCI) energies \cite{QP2} and we also report the MP2 curve and its third-order variant (MP3), which improves upon MP2 towards FCI.
RPA@HF and RPA@{\GOWO}@HF yield almost identical results, and both significantly overestimate the FCI correlation energy, while RPAx@HF and BSE@{\GOWO}@HF slightly over- and undershoot the FCI energy, respectively, RPAx@HF yielding the best match to FCI in the case of \ce{H2}.
Interestingly, the BSE@{\GOWO}@HF scheme yields a more accurate equilibrium bond length than any other method irrespectively of the basis set (see Table in the {\SI}).
For example, BSE@{\GOWO}@HF/cc-pVQZ is only off by $0.003$ bohr as compared to FCI/cc-pVQZ, while RPAx@HF, MP2, and CC2 underestimate the bond length by $0.008$, $0.011$, and $0.011$ bohr, respectively.
The RPA-based schemes are much less accurate, with even shorter equilibrium bond lengths.
This is a general trend that is magnified in larger systems as the ones discussed below.

Despite the shallow nature of its PES, the scenario is almost identical for \ce{LiH} for which we report the CC2, CCSD and CC3 energies in addition to MP2 energies.
In this case, RPAx@HF and BSE@{\GOWO}@HF nestle the CCSD and CC3 energy curves, theses surfaces running almost perfectly parallel to one another.
Here again, the BSE@{\GOWO}@HF/cc-pVQZ equilibrium bond length is extremely accurate ($3.017$ bohr) as compared to CC3/cc-pVQZ ($3.019$ bohr).

The cases of \ce{LiF} and \ce{HCl} (see Fig.~\ref{fig:PES-LiF-HCl}) are chemically interesting as they correspond to strongly polarized bonds towards the halogen atoms which are much more electronegative than the first-column elements.
For these partially ionic bonds, the performance of BSE@{\GOWO}@HF is terrific with an almost perfect match to the CC3 curve.
Maybe surprisingly, BSE@{\GOWO}@HF is on par with both CC2 and CCSD, and outperforms RPAx@HF by a big margin, the latter fact being also observed for the other diatomics discussed below.
Interestingly, while CCSD and CC2 systematically underestimates the total energy, the BSE@{\GOWO}@HF energy is always lower than the reference CC3 energy.
This observation is not only true for \ce{LiF} and \ce{HCl}, but holds for every single systems that is considered herein. 
Moreover, this is consistent with the study by Maggio and Kresse on the HEG showing that BSE slightly overestimates the correlation energy as compared to QMC reference data. \cite{Maggio_2016} 
Similarly, the much larger overestimation of the correlation energy that we observe at the RPA@$GW$ level was also observed for the HEG. 
Care must be taken however in drawing comparisons since the HEG study of Ref.~\onlinecite{Maggio_2016} was performed starting with LDA eigenstates.

For \ce{HCl}, the data reported in Table \ref{tab:Req} show that the BSE@{\GOWO}@HF equilibrium bond length is again in very good agreement with its CC3 counterpart as it underestimates the bond lengths by a few hundredths of bohr only.
However, in the case of \ce{LiF}, the attentive reader can observe a small ``glitch'' in the $GW$-based curves very close to their minimum.
As observed in Refs.~\onlinecite{vanSetten_2015,Maggio_2017,Loos_2018} and explained in details in Refs.~\onlinecite{Veril_2018,Duchemin_2020}, these irregularities, which makes particularly tricky the location of the minima, are due to ``jumps'' between distinct solutions of the $GW$ quasiparticle equation.
Including a broadening via an increase of the $\eta$ value entering in the expression of the $GW$ self-energy and the screened Coulomb operator softens the problem, but does not remove it completely.
When irregularities are present in the PES, we have fitted a Morse potential of the form $M(R) = D_0\qty{1-\exp[-\alpha\qty(R-\Req)]}^2$ to the PES in order to provide an estimate of the equilibrium bond length.
These values are reported in parenthesis in Table \ref{tab:Req}.
For the smooth PES where one can obtain both the genuine minimum and the fitted minimum (\ie, based on the Morse curve), this procedure has been shown to be very accurate with an error of the order of $10^{-3}$ bohr in most cases. 
We note that these irregularities are much smaller than the differences between the BSE and the other RPA-like techniques (RPA, RPAx, RPA@$GW$) leaving BSE unambiguously more accurate than these approaches.

Let us now look at the isoelectronic series \ce{N2}, \ce{CO}, and \ce{BF}, which have a decreasing bond order (from triple to single bond).
The conclusions drawn for the previous systems also apply to these molecules. 
In particular, as shown in Fig.~\ref{fig:PES-N2-CO-BF}, the performance of BSE@{\GOWO}@HF is outstanding with an error of the order of $1\%$ on the correlation energy.
Importantly, it systematically outperforms both CC2 and CCSD.
One can notice some irregularities in the PES of \ce{BF} with the cc-pVDZ et cc-pVTZ basis sets (see the {\SI}).
The PES of \ce{N2} and \ce{CO} are smooth though, and yield accurate equilibrium bond lengths once more. 
Indeed, at the BSE@{\GOWO}@HF/cc-pVQZ level of theory, we obtain $2.065$, $2.134$, and $2.385$ bohr for \ce{N2}, \ce{CO}, and \ce{BF}, respectively, which has to be compared with the CC3/cc-pVQZ values of $2.075$, $2.136$ and $2.390$ bohr, respectively.

As a final example, we consider the \ce{F2} molecule, a notoriously difficult case to treat due to the weakness of its covalent bond (see Fig.~\ref{fig:PES-F2}), hence its relatively long equilibrium bond length ($2.663$ bohr at the CC3/cc-pVQZ level).
Similarly to what is observed for \ce{LiF} and \ce{BF}, there are irregularities near the minimum of the {\GOWO}-based curves.
However, BSE@{\GOWO}@HF is the closest to the CC3 curve, with an error on the correlation energy of $1\%$ and an estimated bond length of $2.640$ bohr (via a Morse fit) at the BSE@{\GOWO}@HF/cc-pVQZ level.
Note that, for this system, triplet (and then singlet) instabilities appear for quite short bond lengths.
However, around the equilibrium structure, we have not encountered any instabilities. 
This is an important outcome of the present study as the difficulties encountered at large interatomic distances (\ie, close to the dissociation limit) do not prevent the BSE approach to be potentially useful and accurate in the vicinity of equilibrium distances. 
Furthermore, preliminary calculations could not detect any singlet instabilities in the vicinity of the lowest singlet excited-state minima.

As a final remark, we would like to mention that although we have considered here only a limited set of compounds, our correlation energy mean absolute error (MAE) with BSE@{\GOWO}@HF of \titou{$4.7$} mHa (as compared to CC3) is significantly smaller than the one obtained with MP2, CC2, and CCSD ($18.2$, $13.1$ and $13.5$ mHa respectively). 
For comparison, the RPA-related formalisms return larger MAEs of $75.6$, $43.1$, and $68.2$ mHa for BSE@{\GOWO}@HF, RPAx@HF, and RPA@HF, respectively.

In this Letter, we hope to have illustrated that the ACFDT@BSE formalism is a promising methodology for the computation of accurate ground-state PES and their corresponding equilibrium structures.
To do so, we have shown that calculating the BSE correlation energy computed within the ACFDT framework yields extremely accurate PES around equilibrium.
\titou{Their accuracy near the dissociation limit remains an open question. \cite{Caruso_2013,Olsen_2014,Colonna_2014,Hellgren_2015,Holzer_2018}}
We have illustrated this for 8 diatomic molecules for which we have also computed reference ground-state energies using coupled cluster methods (CC2, CCSD, and CC3).
Moreover, because triplet states do not contribute to the ACFDT correlation energy and singlet instabilities do not appear for weakly-correlated systems around their equilibrium structure, the present scheme does not suffer from singlet nor triplet instabilities.
However, we have also observed that, in some cases, unphysical irregularities on the ground-state PES due to the appearance of discontinuities as a function of the bond length for some of the $GW$ quasiparticle energies. 
Such an unphysical behaviour stems from defining the quasiparticle energy as the solution of the quasiparticle equation with the largest spectral weight in cases where several solutions can be found.
This shortcoming has been thoroughly described in several previous studies.\cite{vanSetten_2015,Maggio_2017,Loos_2018,Veril_2018,Duchemin_2020}
We believe that this central issue must be resolved if one wants to expand the applicability of the present method.

\section*{Acknowledgements}
PFL would like to thank Julien Toulouse for enlightening discussions about RPA, and XB is indebted to Valerio Olevano for numerous discussions.
This work was performed using HPC resources from GENCI-TGCC (Grant No.~2019-A0060801738), CALMIP (Toulouse) under allocation 2020-18005, and the CCIPL center installed in Nantes.
Funding from the \textit{``Centre National de la Recherche Scientifique''} is acknowledged.
This work has also been supported through the EUR grant NanoX ANR-17-EURE-0009 in the framework of the \textit{``Programme des Investissements d'Avenir''.} 

\section*{Supporting Information}
See {\SI} for additional potential energy curves computed with other basis sets and within the frozen-core approximation, tables gathering equilibrium distances for smaller basis sets (cc-pVDZ and cc-pVTZ), \titou{CC3, CCSDT, and CCSDT(Q) total energies, as well as comparisons between the extended and regular BSE schemes (correlation energies at equilibrium geometries).}

\bibliography{BSE-PES,BSE-PES-control}

\providecommand{\latin}[1]{#1}
\makeatletter
\providecommand{\doi}
  {\begingroup\let\do\@makeother\dospecials
  \catcode`\{=1 \catcode`\}=2 \doi@aux}
\providecommand{\doi@aux}[1]{\endgroup\texttt{#1}}
\makeatother
\providecommand*\mcitethebibliography{\thebibliography}
\csname @ifundefined\endcsname{endmcitethebibliography}
  {\let\endmcitethebibliography\endthebibliography}{}
\begin{mcitethebibliography}{107}
\providecommand*\natexlab[1]{#1}
\providecommand*\mciteSetBstSublistMode[1]{}
\providecommand*\mciteSetBstMaxWidthForm[2]{}
\providecommand*\mciteBstWouldAddEndPuncttrue
  {\def\EndOfBibitem{\unskip.}}
\providecommand*\mciteBstWouldAddEndPunctfalse
  {\let\EndOfBibitem\relax}
\providecommand*\mciteSetBstMidEndSepPunct[3]{}
\providecommand*\mciteSetBstSublistLabelBeginEnd[3]{}
\providecommand*\EndOfBibitem{}
\mciteSetBstSublistMode{f}
\mciteSetBstMaxWidthForm{subitem}{(\alph{mcitesubitemcount})}
\mciteSetBstSublistLabelBeginEnd
  {\mcitemaxwidthsubitemform\space}
  {\relax}
  {\relax}

\bibitem[Runge and Gross(1984)Runge, and Gross]{Runge_1984}
Runge,~E.; Gross,~E. K.~U. Density-Functional Theory for Time-Dependent
  Systems. \emph{Phys. Rev. Lett.} \textbf{1984}, \emph{52}, 997--1000\relax
\mciteBstWouldAddEndPuncttrue
\mciteSetBstMidEndSepPunct{\mcitedefaultmidpunct}
{\mcitedefaultendpunct}{\mcitedefaultseppunct}\relax
\EndOfBibitem
\bibitem[Casida(1995)]{Casida}
Casida,~M.~E. In \emph{Time-Dependent Density Functional Response Theory for
  Molecules}; Chong,~D.~P., Ed.; Recent Advances in Density Functional Methods;
  World Scientific, Singapore, 1995; pp 155--192\relax
\mciteBstWouldAddEndPuncttrue
\mciteSetBstMidEndSepPunct{\mcitedefaultmidpunct}
{\mcitedefaultendpunct}{\mcitedefaultseppunct}\relax
\EndOfBibitem
\bibitem[Salpeter and Bethe(1951)Salpeter, and Bethe]{Salpeter_1951}
Salpeter,~E.~E.; Bethe,~H.~A. A Relativistic Equation for Bound-State Problems.
  \emph{Phys. Rev.} \textbf{1951}, \emph{84}, 1232\relax
\mciteBstWouldAddEndPuncttrue
\mciteSetBstMidEndSepPunct{\mcitedefaultmidpunct}
{\mcitedefaultendpunct}{\mcitedefaultseppunct}\relax
\EndOfBibitem
\bibitem[Strinati(1988)]{Strinati_1988}
Strinati,~G. Application of the {{Green}}'s Functions Method to the Study of
  the Optical Properties of Semiconductors. \emph{Riv. Nuovo Cimento}
  \textbf{1988}, \emph{11}, 1--86\relax
\mciteBstWouldAddEndPuncttrue
\mciteSetBstMidEndSepPunct{\mcitedefaultmidpunct}
{\mcitedefaultendpunct}{\mcitedefaultseppunct}\relax
\EndOfBibitem
\bibitem[Albrecht \latin{et~al.}(1998)Albrecht, Reining, Del~Sole, and
  Onida]{Albrecht_1998}
Albrecht,~S.; Reining,~L.; Del~Sole,~R.; Onida,~G. Ab Initio Calculation of
  Excitonic Effects in the Optical Spectra of Semiconductors. \emph{Phys. Rev.
  Lett.} \textbf{1998}, \emph{80}, 4510--4513\relax
\mciteBstWouldAddEndPuncttrue
\mciteSetBstMidEndSepPunct{\mcitedefaultmidpunct}
{\mcitedefaultendpunct}{\mcitedefaultseppunct}\relax
\EndOfBibitem
\bibitem[Rohlfing and Louie(1998)Rohlfing, and Louie]{Rohlfing_1998}
Rohlfing,~M.; Louie,~S.~G. Electron-Hole Excitations in Semiconductors and
  Insulators. \emph{Phys. Rev. Lett.} \textbf{1998}, \emph{81},
  2312--2315\relax
\mciteBstWouldAddEndPuncttrue
\mciteSetBstMidEndSepPunct{\mcitedefaultmidpunct}
{\mcitedefaultendpunct}{\mcitedefaultseppunct}\relax
\EndOfBibitem
\bibitem[Benedict \latin{et~al.}(1998)Benedict, Shirley, and
  Bohn]{Benedict_1998}
Benedict,~L.~X.; Shirley,~E.~L.; Bohn,~R.~B. Optical Absorption of Insulators
  and the Electron-Hole Interaction: An Ab Initio Calculation. \emph{Phys. Rev.
  Lett.} \textbf{1998}, \emph{80}, 4514--4517\relax
\mciteBstWouldAddEndPuncttrue
\mciteSetBstMidEndSepPunct{\mcitedefaultmidpunct}
{\mcitedefaultendpunct}{\mcitedefaultseppunct}\relax
\EndOfBibitem
\bibitem[van~der Horst \latin{et~al.}(1999)van~der Horst, Bobbert, Michels,
  Brocks, and Kelly]{vanderHorst_1999}
van~der Horst,~J.-W.; Bobbert,~P.~A.; Michels,~M. A.~J.; Brocks,~G.;
  Kelly,~P.~J. Ab Initio Calculation of the Electronic and Optical Excitations
  in Polythiophene: Effects of Intra- and Interchain Screening. \emph{Phys.
  Rev. Lett.} \textbf{1999}, \emph{83}, 4413--4416\relax
\mciteBstWouldAddEndPuncttrue
\mciteSetBstMidEndSepPunct{\mcitedefaultmidpunct}
{\mcitedefaultendpunct}{\mcitedefaultseppunct}\relax
\EndOfBibitem
\bibitem[Ma \latin{et~al.}(2009)Ma, Rohlfing, and Molteni]{Ma_2009}
Ma,~Y.; Rohlfing,~M.; Molteni,~C. Excited States of Biological Chromophores
  Studied Using Many-Body Perturbation Theory: Effects of Resonant-Antiresonant
  Coupling and Dynamical Screening. \emph{Phys. Rev. B} \textbf{2009},
  \emph{80}, 241405\relax
\mciteBstWouldAddEndPuncttrue
\mciteSetBstMidEndSepPunct{\mcitedefaultmidpunct}
{\mcitedefaultendpunct}{\mcitedefaultseppunct}\relax
\EndOfBibitem
\bibitem[Puschnig and Ambrosch-Draxl(2002)Puschnig, and
  Ambrosch-Draxl]{Pushchnig_2002}
Puschnig,~P.; Ambrosch-Draxl,~C. Suppression of Electron-Hole Correlations in
  3D Polymer Materials. \emph{Phys. Rev. Lett.} \textbf{2002}, \emph{89},
  056405\relax
\mciteBstWouldAddEndPuncttrue
\mciteSetBstMidEndSepPunct{\mcitedefaultmidpunct}
{\mcitedefaultendpunct}{\mcitedefaultseppunct}\relax
\EndOfBibitem
\bibitem[Tiago \latin{et~al.}(2003)Tiago, Northrup, and Louie]{Tiago_2003}
Tiago,~M.~L.; Northrup,~J.~E.; Louie,~S.~G. Ab Initio Calculation of the
  Electronic and Optical Properties of Solid Pentacene. \emph{Phys. Rev. B}
  \textbf{2003}, \emph{67}, 115212\relax
\mciteBstWouldAddEndPuncttrue
\mciteSetBstMidEndSepPunct{\mcitedefaultmidpunct}
{\mcitedefaultendpunct}{\mcitedefaultseppunct}\relax
\EndOfBibitem
\bibitem[Palummo \latin{et~al.}(2009)Palummo, Hogan, Sottile, Bagal\'{a}, and
  Rubio]{Palumno_2009}
Palummo,~M.; Hogan,~C.; Sottile,~F.; Bagal\'{a},~P.; Rubio,~A. Ab Initio
  Electronic and Optical Spectra of Free-Base Porphyrins: The Role of
  Electronic Correlation. \emph{J. Chem. Phys.} \textbf{2009}, \emph{131},
  084102\relax
\mciteBstWouldAddEndPuncttrue
\mciteSetBstMidEndSepPunct{\mcitedefaultmidpunct}
{\mcitedefaultendpunct}{\mcitedefaultseppunct}\relax
\EndOfBibitem
\bibitem[Rocca \latin{et~al.}(2010)Rocca, Lu, and Galli]{Rocca_2010}
Rocca,~D.; Lu,~D.; Galli,~G. Ab Initio Calculations of Optical Absorption
  Spectra: Solution of the Bethe--Salpeter Equation Within Density Matrix
  Perturbation Theory. \emph{J. Chem. Phys.} \textbf{2010}, \emph{133},
  164109\relax
\mciteBstWouldAddEndPuncttrue
\mciteSetBstMidEndSepPunct{\mcitedefaultmidpunct}
{\mcitedefaultendpunct}{\mcitedefaultseppunct}\relax
\EndOfBibitem
\bibitem[Sharifzadeh \latin{et~al.}(2012)Sharifzadeh, Biller, Kronik, and
  Neaton]{Sharifzadeh_2012}
Sharifzadeh,~S.; Biller,~A.; Kronik,~L.; Neaton,~J.~B. Quasiparticle and
  Optical Spectroscopy of the Organic Semiconductors Pentacene and PTCDA From
  First Principles. \emph{Phys. Rev. B} \textbf{2012}, \emph{85}, 125307\relax
\mciteBstWouldAddEndPuncttrue
\mciteSetBstMidEndSepPunct{\mcitedefaultmidpunct}
{\mcitedefaultendpunct}{\mcitedefaultseppunct}\relax
\EndOfBibitem
\bibitem[Cudazzo \latin{et~al.}(2012)Cudazzo, Gatti, and Rubio]{Cudazzo_2012}
Cudazzo,~P.; Gatti,~M.; Rubio,~A. Excitons in Molecular Crystals From
  First-Principles Many-Body Perturbation Theory: Picene Versus Pentacene.
  \emph{Phys. Rev. B} \textbf{2012}, \emph{86}, 195307\relax
\mciteBstWouldAddEndPuncttrue
\mciteSetBstMidEndSepPunct{\mcitedefaultmidpunct}
{\mcitedefaultendpunct}{\mcitedefaultseppunct}\relax
\EndOfBibitem
\bibitem[Boulanger \latin{et~al.}(2014)Boulanger, Jacquemin, Duchemin, and
  Blase]{Boulanger_2014}
Boulanger,~P.; Jacquemin,~D.; Duchemin,~I.; Blase,~X. Fast and {{Accurate
  Electronic Excitations}} in {{Cyanines}} with the {{Many}}-{{Body
  Bethe}}\textendash{}{{Salpeter Approach}}. \emph{J. Chem. Theory Comput.}
  \textbf{2014}, \emph{10}, 1212--1218\relax
\mciteBstWouldAddEndPuncttrue
\mciteSetBstMidEndSepPunct{\mcitedefaultmidpunct}
{\mcitedefaultendpunct}{\mcitedefaultseppunct}\relax
\EndOfBibitem
\bibitem[Ljungberg \latin{et~al.}(2015)Ljungberg, Koval, Ferrari, Foerster, and
  S\'anchez-Portal]{Ljungberg_2015}
Ljungberg,~M.~P.; Koval,~P.; Ferrari,~F.; Foerster,~D.; S\'anchez-Portal,~D.
  Cubic-Scaling Iterative Solution of the Bethe-Salpeter Equation for Finite
  Systems. \emph{Phys. Rev. B} \textbf{2015}, \emph{92}, 075422\relax
\mciteBstWouldAddEndPuncttrue
\mciteSetBstMidEndSepPunct{\mcitedefaultmidpunct}
{\mcitedefaultendpunct}{\mcitedefaultseppunct}\relax
\EndOfBibitem
\bibitem[Hirose \latin{et~al.}(2015)Hirose, Noguchi, and Sugino]{Hirose_2015}
Hirose,~D.; Noguchi,~Y.; Sugino,~O. All-Electron GW+Bethe-Salpeter Calculations
  on Small Molecules. \emph{Phys. Rev. B} \textbf{2015}, \emph{91},
  205111\relax
\mciteBstWouldAddEndPuncttrue
\mciteSetBstMidEndSepPunct{\mcitedefaultmidpunct}
{\mcitedefaultendpunct}{\mcitedefaultseppunct}\relax
\EndOfBibitem
\bibitem[Cocchi and Draxl(2015)Cocchi, and Draxl]{Cocchi_2015}
Cocchi,~C.; Draxl,~C. Bound Excitons and Many-Body Effects in X-Ray Absorption
  Spectra of Azobenzene-Functionalized Self-Assembled Monolayers. \emph{Phys.
  Rev. B} \textbf{2015}, \emph{92}, 205105\relax
\mciteBstWouldAddEndPuncttrue
\mciteSetBstMidEndSepPunct{\mcitedefaultmidpunct}
{\mcitedefaultendpunct}{\mcitedefaultseppunct}\relax
\EndOfBibitem
\bibitem[Ziaei and Bredow(2017)Ziaei, and Bredow]{Ziaei_2017}
Ziaei,~V.; Bredow,~T. Simple many-body based screening mixing ansatz for
  improvement of $GW$/Bethe-Salpeter equation excitation energies of molecular
  systems. \emph{Phys. Rev. B} \textbf{2017}, \emph{96}, 195115\relax
\mciteBstWouldAddEndPuncttrue
\mciteSetBstMidEndSepPunct{\mcitedefaultmidpunct}
{\mcitedefaultendpunct}{\mcitedefaultseppunct}\relax
\EndOfBibitem
\bibitem[Refaely-Abramson \latin{et~al.}(2017)Refaely-Abramson, da~Jornada,
  Louie, and Neaton]{Abramson_2017}
Refaely-Abramson,~S.; da~Jornada,~F.~H.; Louie,~S.~G.; Neaton,~J.~B. Origins of
  Singlet Fission in Solid Pentacene from an ab initio Green's Function
  Approach. \emph{Phys. Rev. Lett.} \textbf{2017}, \emph{119}, 267401\relax
\mciteBstWouldAddEndPuncttrue
\mciteSetBstMidEndSepPunct{\mcitedefaultmidpunct}
{\mcitedefaultendpunct}{\mcitedefaultseppunct}\relax
\EndOfBibitem
\bibitem[Gonz{\'a}lez \latin{et~al.}(2012)Gonz{\'a}lez, Escudero, and
  Serrano-Andr\`es]{Gonzales_2012}
Gonz{\'a}lez,~L.; Escudero,~D.; Serrano-Andr\`es,~L. Progress and Challenges in
  the Calculation of Electronic Excited States. \emph{ChemPhysChem}
  \textbf{2012}, \emph{13}, 28--51\relax
\mciteBstWouldAddEndPuncttrue
\mciteSetBstMidEndSepPunct{\mcitedefaultmidpunct}
{\mcitedefaultendpunct}{\mcitedefaultseppunct}\relax
\EndOfBibitem
\bibitem[Loos \latin{et~al.}(2020)Loos, Scemama, and Jacquemin]{Loos_2020a}
Loos,~P.~F.; Scemama,~A.; Jacquemin,~D. The Quest for Highly-Accurate
  Excitation Energies: a Computational Perspective. \emph{J. Phys. Chem. Lett.}
  \textbf{2020}, submitted\relax
\mciteBstWouldAddEndPuncttrue
\mciteSetBstMidEndSepPunct{\mcitedefaultmidpunct}
{\mcitedefaultendpunct}{\mcitedefaultseppunct}\relax
\EndOfBibitem
\bibitem[Jacquemin \latin{et~al.}(2015)Jacquemin, Duchemin, and
  Blase]{Jacquemin_2015}
Jacquemin,~D.; Duchemin,~I.; Blase,~X. Benchmarking the
  {{Bethe}}\textendash{}{{Salpeter Formalism}} on a {{Standard Organic
  Molecular Set}}. \emph{J. Chem. Theory Comput.} \textbf{2015}, \emph{11},
  3290--3304\relax
\mciteBstWouldAddEndPuncttrue
\mciteSetBstMidEndSepPunct{\mcitedefaultmidpunct}
{\mcitedefaultendpunct}{\mcitedefaultseppunct}\relax
\EndOfBibitem
\bibitem[Bruneval \latin{et~al.}(2015)Bruneval, Hamed, and
  Neaton]{Bruneval_2015}
Bruneval,~F.; Hamed,~S.~M.; Neaton,~J.~B. A Systematic Benchmark of the
  {\emph{Ab Initio}} {{Bethe}}-{{Salpeter}} Equation Approach for Low-Lying
  Optical Excitations of Small Organic Molecules. \emph{J. Chem. Phys.}
  \textbf{2015}, \emph{142}, 244101\relax
\mciteBstWouldAddEndPuncttrue
\mciteSetBstMidEndSepPunct{\mcitedefaultmidpunct}
{\mcitedefaultendpunct}{\mcitedefaultseppunct}\relax
\EndOfBibitem
\bibitem[Hung \latin{et~al.}(2016)Hung, {da Jornada}, Souto-Casares,
  Chelikowsky, Louie, and Ogut]{Hung_2016}
Hung,~L.; {da Jornada},~F.~H.; Souto-Casares,~J.; Chelikowsky,~J.~R.;
  Louie,~S.~G.; Ogut,~S. Excitation Spectra of Aromatic Molecules within a
  Real-Space {{GW}}-{{BSE}} Formalism: {{Role}} of Self-Consistency and Vertex
  Corrections. \emph{Phys. Rev. B} \textbf{2016}, \emph{94}, 085125\relax
\mciteBstWouldAddEndPuncttrue
\mciteSetBstMidEndSepPunct{\mcitedefaultmidpunct}
{\mcitedefaultendpunct}{\mcitedefaultseppunct}\relax
\EndOfBibitem
\bibitem[Hung \latin{et~al.}(2017)Hung, Bruneval, Baishya, and {\"O}{\u
  g}{\"u}t]{Hung_2017}
Hung,~L.; Bruneval,~F.; Baishya,~K.; {\"O}{\u g}{\"u}t,~S. Benchmarking the
  {{{\emph{GW}}}} {{Approximation}} and {{Bethe}}\textendash{}{{Salpeter
  Equation}} for {{Groups IB}} and {{IIB Atoms}} and {{Monoxides}}. \emph{J.
  Chem. Theory Comput.} \textbf{2017}, \emph{13}, 2135--2146\relax
\mciteBstWouldAddEndPuncttrue
\mciteSetBstMidEndSepPunct{\mcitedefaultmidpunct}
{\mcitedefaultendpunct}{\mcitedefaultseppunct}\relax
\EndOfBibitem
\bibitem[Krause and Klopper(2017)Krause, and Klopper]{Krause_2017}
Krause,~K.; Klopper,~W. Implementation of the {{Bethe}}-{{Salpeter}} Equation
  in the {{Turbomole}} Program. \emph{J. Comput. Chem.} \textbf{2017},
  \emph{38}, 383--388\relax
\mciteBstWouldAddEndPuncttrue
\mciteSetBstMidEndSepPunct{\mcitedefaultmidpunct}
{\mcitedefaultendpunct}{\mcitedefaultseppunct}\relax
\EndOfBibitem
\bibitem[Jacquemin \latin{et~al.}(2017)Jacquemin, Duchemin, Blondel, and
  Blase]{Jacquemin_2017}
Jacquemin,~D.; Duchemin,~I.; Blondel,~A.; Blase,~X. Benchmark of
  {{Bethe}}-{{Salpeter}} for {{Triplet Excited}}-{{States}}. \emph{J. Chem.
  Theory Comput.} \textbf{2017}, \emph{13}, 767--783\relax
\mciteBstWouldAddEndPuncttrue
\mciteSetBstMidEndSepPunct{\mcitedefaultmidpunct}
{\mcitedefaultendpunct}{\mcitedefaultseppunct}\relax
\EndOfBibitem
\bibitem[Blase \latin{et~al.}(2018)Blase, Duchemin, and Jacquemin]{Blase_2018}
Blase,~X.; Duchemin,~I.; Jacquemin,~D. The {{Bethe}}\textendash{}{{Salpeter}}
  Equation in Chemistry: Relations with {{TD}}-{{DFT}}, Applications and
  Challenges. \emph{Chem. Soc. Rev.} \textbf{2018}, \emph{47}, 1022--1043\relax
\mciteBstWouldAddEndPuncttrue
\mciteSetBstMidEndSepPunct{\mcitedefaultmidpunct}
{\mcitedefaultendpunct}{\mcitedefaultseppunct}\relax
\EndOfBibitem
\bibitem[Garcia-Lastra and Thygesen(2011)Garcia-Lastra, and
  Thygesen]{Lastra_2011}
Garcia-Lastra,~J.~M.; Thygesen,~K.~S. Renormalization of Optical Excitations in
  Molecules Near a Metal Surface. \emph{Phys. Rev. Lett.} \textbf{2011},
  \emph{106}, 187402\relax
\mciteBstWouldAddEndPuncttrue
\mciteSetBstMidEndSepPunct{\mcitedefaultmidpunct}
{\mcitedefaultendpunct}{\mcitedefaultseppunct}\relax
\EndOfBibitem
\bibitem[Blase and Attaccalite(2011)Blase, and Attaccalite]{Blase_2011b}
Blase,~X.; Attaccalite,~C. Charge-Transfer Excitations in Molecular
  Donor-Acceptor Complexes Within the Many-Body Bethe-Salpeter Approach.
  \emph{Appl. Phys. Lett.} \textbf{2011}, \emph{99}, 171909\relax
\mciteBstWouldAddEndPuncttrue
\mciteSetBstMidEndSepPunct{\mcitedefaultmidpunct}
{\mcitedefaultendpunct}{\mcitedefaultseppunct}\relax
\EndOfBibitem
\bibitem[Baumeier \latin{et~al.}(2012)Baumeier, Andrienko, and
  Rohlfing]{Baumeier_2012}
Baumeier,~B.; Andrienko,~D.; Rohlfing,~M. Frenkel and Charge-Transfer
  Excitations in Donor--Acceptor Complexes From Many-Body Green's Functions
  Theory. \emph{J. Chem. Theory Comput.} \textbf{2012}, \emph{8},
  2790--2795\relax
\mciteBstWouldAddEndPuncttrue
\mciteSetBstMidEndSepPunct{\mcitedefaultmidpunct}
{\mcitedefaultendpunct}{\mcitedefaultseppunct}\relax
\EndOfBibitem
\bibitem[Duchemin \latin{et~al.}(2012)Duchemin, Deutsch, and
  Blase]{Duchemin_2012}
Duchemin,~I.; Deutsch,~T.; Blase,~X. Short-Range to Long-Range Charge-Transfer
  Excitations in the Zincbacteriochlorin-Bacteriochlorin Complex: A
  Bethe-Salpeter Study. \emph{Phys. Rev. Lett.} \textbf{2012}, \emph{109},
  167801\relax
\mciteBstWouldAddEndPuncttrue
\mciteSetBstMidEndSepPunct{\mcitedefaultmidpunct}
{\mcitedefaultendpunct}{\mcitedefaultseppunct}\relax
\EndOfBibitem
\bibitem[Cudazzo \latin{et~al.}(2013)Cudazzo, Gatti, Rubio, and
  Sottile]{Cudazzo_2013}
Cudazzo,~P.; Gatti,~M.; Rubio,~A.; Sottile,~F. Frenkel Versus Charge-Transfer
  Exciton Dispersion in Molecular Crystals. \emph{Phys. Rev. B} \textbf{2013},
  \emph{88}, 195152\relax
\mciteBstWouldAddEndPuncttrue
\mciteSetBstMidEndSepPunct{\mcitedefaultmidpunct}
{\mcitedefaultendpunct}{\mcitedefaultseppunct}\relax
\EndOfBibitem
\bibitem[Ziaei and Bredow(2016)Ziaei, and Bredow]{Ziaei_2016}
Ziaei,~V.; Bredow,~T. GW-BSE Approach on S1 Vertical Transition Energy of Large
  Charge Transfer Compounds: A Performance Assessment. \emph{J. Chem. Phys.}
  \textbf{2016}, \emph{145}, 174305\relax
\mciteBstWouldAddEndPuncttrue
\mciteSetBstMidEndSepPunct{\mcitedefaultmidpunct}
{\mcitedefaultendpunct}{\mcitedefaultseppunct}\relax
\EndOfBibitem
\bibitem[Hybertsen and Louie(1986)Hybertsen, and Louie]{Hybertsen_1986}
Hybertsen,~M.~S.; Louie,~S.~G. Electron Correlation in Semiconductors and
  Insulators: {{Band}} Gaps and Quasiparticle Energies. \emph{Phys. Rev. B}
  \textbf{1986}, \emph{34}, 5390--5413\relax
\mciteBstWouldAddEndPuncttrue
\mciteSetBstMidEndSepPunct{\mcitedefaultmidpunct}
{\mcitedefaultendpunct}{\mcitedefaultseppunct}\relax
\EndOfBibitem
\bibitem[Shishkin and Kresse(2007)Shishkin, and Kresse]{Shishkin_2007}
Shishkin,~M.; Kresse,~G. Self-Consistent {{G W}} Calculations for
  Semiconductors and Insulators. \emph{Phys. Rev. B} \textbf{2007}, \emph{75},
  235102\relax
\mciteBstWouldAddEndPuncttrue
\mciteSetBstMidEndSepPunct{\mcitedefaultmidpunct}
{\mcitedefaultendpunct}{\mcitedefaultseppunct}\relax
\EndOfBibitem
\bibitem[Blase \latin{et~al.}(2011)Blase, Attaccalite, and Olevano]{Blase_2011}
Blase,~X.; Attaccalite,~C.; Olevano,~V. First-Principles {{GW}} Calculations
  for Fullerenes, Porphyrins, Phtalocyanine, and Other Molecules of Interest
  for Organic Photovoltaic Applications. \emph{Phys. Rev. B} \textbf{2011},
  \emph{83}, 115103\relax
\mciteBstWouldAddEndPuncttrue
\mciteSetBstMidEndSepPunct{\mcitedefaultmidpunct}
{\mcitedefaultendpunct}{\mcitedefaultseppunct}\relax
\EndOfBibitem
\bibitem[Faber \latin{et~al.}(2011)Faber, Attaccalite, Olevano, Runge, and
  Blase]{Faber_2011}
Faber,~C.; Attaccalite,~C.; Olevano,~V.; Runge,~E.; Blase,~X. First-Principles
  {{GW}} Calculations for {{DNA}} and {{RNA}} Nucleobases. \emph{Phys. Rev. B}
  \textbf{2011}, \emph{83}, 115123\relax
\mciteBstWouldAddEndPuncttrue
\mciteSetBstMidEndSepPunct{\mcitedefaultmidpunct}
{\mcitedefaultendpunct}{\mcitedefaultseppunct}\relax
\EndOfBibitem
\bibitem[Rangel \latin{et~al.}(2016)Rangel, Hamed, Bruneval, and
  Neaton]{Rangel_2016}
Rangel,~T.; Hamed,~S.~M.; Bruneval,~F.; Neaton,~J.~B. Evaluating the GW
  Approximation with CCSD(T) for Charged Excitations Across the Oligoacenes.
  \emph{J. Chem. Theory Comput.} \textbf{2016}, \emph{12}, 2834--2842\relax
\mciteBstWouldAddEndPuncttrue
\mciteSetBstMidEndSepPunct{\mcitedefaultmidpunct}
{\mcitedefaultendpunct}{\mcitedefaultseppunct}\relax
\EndOfBibitem
\bibitem[Kaplan \latin{et~al.}(2016)Kaplan, Harding, Seiler, Weigend, Evers,
  and {van Setten}]{Kaplan_2016}
Kaplan,~F.; Harding,~M.~E.; Seiler,~C.; Weigend,~F.; Evers,~F.; {van
  Setten},~M.~J. Quasi-{{Particle Self}}-{{Consistent}} {{{\emph{GW}}}} for
  {{Molecules}}. \emph{J. Chem. Theory Comput.} \textbf{2016}, \emph{12},
  2528--2541\relax
\mciteBstWouldAddEndPuncttrue
\mciteSetBstMidEndSepPunct{\mcitedefaultmidpunct}
{\mcitedefaultendpunct}{\mcitedefaultseppunct}\relax
\EndOfBibitem
\bibitem[Gui \latin{et~al.}(2018)Gui, Holzer, and Klopper]{Gui_2018}
Gui,~X.; Holzer,~C.; Klopper,~W. Accuracy {{Assessment}} of {{{\emph{GW}}}}
  {{Starting Points}} for {{Calculating Molecular Excitation Energies Using}}
  the {{Bethe}}\textendash{{Salpeter Formalism}}. \emph{J. Chem. Theory
  Comput.} \textbf{2018}, \emph{14}, 2127--2136\relax
\mciteBstWouldAddEndPuncttrue
\mciteSetBstMidEndSepPunct{\mcitedefaultmidpunct}
{\mcitedefaultendpunct}{\mcitedefaultseppunct}\relax
\EndOfBibitem
\bibitem[Levine \latin{et~al.}(2006)Levine, Ko, Quenneville, and
  Mart\'Inez]{Levine_2006}
Levine,~B.~G.; Ko,~C.; Quenneville,~J.; Mart\'Inez,~T.~J. Conical Intersections
  and Double Excitations in Time-Dependent Density Functional Theory.
  \emph{Mol. Phys.} \textbf{2006}, \emph{104}, 1039--1051\relax
\mciteBstWouldAddEndPuncttrue
\mciteSetBstMidEndSepPunct{\mcitedefaultmidpunct}
{\mcitedefaultendpunct}{\mcitedefaultseppunct}\relax
\EndOfBibitem
\bibitem[Tozer and Handy(2000)Tozer, and Handy]{Tozer_2000}
Tozer,~D.~J.; Handy,~N.~C. On the Determination of Excitation Energies Using
  Density Functional Theory. \emph{Phys. Chem. Chem. Phys.} \textbf{2000},
  \emph{2}, 2117--2121\relax
\mciteBstWouldAddEndPuncttrue
\mciteSetBstMidEndSepPunct{\mcitedefaultmidpunct}
{\mcitedefaultendpunct}{\mcitedefaultseppunct}\relax
\EndOfBibitem
\bibitem[{Huix-Rotllant} \latin{et~al.}(2010){Huix-Rotllant}, Natarajan,
  Ipatov, Muhavini~Wawire, Deutsch, and Casida]{Huix-Rotllant_2010}
{Huix-Rotllant},~M.; Natarajan,~B.; Ipatov,~A.; Muhavini~Wawire,~C.;
  Deutsch,~T.; Casida,~M.~E. Assessment of Noncollinear Spin-Flip
  {{Tamm}}\textendash{{Dancoff}} Approximation Time-Dependent
  Density-Functional Theory for the Photochemical Ring-Opening of Oxirane.
  \emph{Phys. Chem. Chem. Phys.} \textbf{2010}, \emph{12}, 12811\relax
\mciteBstWouldAddEndPuncttrue
\mciteSetBstMidEndSepPunct{\mcitedefaultmidpunct}
{\mcitedefaultendpunct}{\mcitedefaultseppunct}\relax
\EndOfBibitem
\bibitem[Elliott \latin{et~al.}(2011)Elliott, Goldson, Canahui, and
  Maitra]{Elliott_2011}
Elliott,~P.; Goldson,~S.; Canahui,~C.; Maitra,~N.~T. Perspectives on
  Double-Excitations in {{TDDFT}}. \emph{Chem. Phys.} \textbf{2011},
  \emph{391}, 110--119\relax
\mciteBstWouldAddEndPuncttrue
\mciteSetBstMidEndSepPunct{\mcitedefaultmidpunct}
{\mcitedefaultendpunct}{\mcitedefaultseppunct}\relax
\EndOfBibitem
\bibitem[Romaniello \latin{et~al.}(2009)Romaniello, Sangalli, Berger, Sottile,
  Molinari, Reining, and Onida]{Romaniello_2009a}
Romaniello,~P.; Sangalli,~D.; Berger,~J.~A.; Sottile,~F.; Molinari,~L.~G.;
  Reining,~L.; Onida,~G. Double Excitations in Finite Systems. \emph{J. Chem.
  Phys.} \textbf{2009}, \emph{130}, 044108\relax
\mciteBstWouldAddEndPuncttrue
\mciteSetBstMidEndSepPunct{\mcitedefaultmidpunct}
{\mcitedefaultendpunct}{\mcitedefaultseppunct}\relax
\EndOfBibitem
\bibitem[Sangalli \latin{et~al.}(2011)Sangalli, Romaniello, Onida, and
  Marini]{Sangalli_2011}
Sangalli,~D.; Romaniello,~P.; Onida,~G.; Marini,~A. Double Excitations in
  Correlated Systems: A Many--Body Approach. \emph{J. Chem. Phys.}
  \textbf{2011}, \emph{134}, 034115\relax
\mciteBstWouldAddEndPuncttrue
\mciteSetBstMidEndSepPunct{\mcitedefaultmidpunct}
{\mcitedefaultendpunct}{\mcitedefaultseppunct}\relax
\EndOfBibitem
\bibitem[Loos \latin{et~al.}(2019)Loos, Boggio-Pasqua, Scemama, Caffarel, and
  Jacquemin]{Loos_2019}
Loos,~P.-F.; Boggio-Pasqua,~M.; Scemama,~A.; Caffarel,~M.; Jacquemin,~D.
  Reference Energies for Double Excitations. \emph{J. Chem. Theory Comput.}
  \textbf{2019}, \emph{15}, 1939--1956\relax
\mciteBstWouldAddEndPuncttrue
\mciteSetBstMidEndSepPunct{\mcitedefaultmidpunct}
{\mcitedefaultendpunct}{\mcitedefaultseppunct}\relax
\EndOfBibitem
\bibitem[Furche and Ahlrichs(2002)Furche, and Ahlrichs]{Furche_2002}
Furche,~F.; Ahlrichs,~R. Adiabatic Time-Dependent Density Functional Methods
  for Excited State Properties. \emph{J. Chem. Phys.} \textbf{2002},
  \emph{117}, 7433\relax
\mciteBstWouldAddEndPuncttrue
\mciteSetBstMidEndSepPunct{\mcitedefaultmidpunct}
{\mcitedefaultendpunct}{\mcitedefaultseppunct}\relax
\EndOfBibitem
\bibitem[Bernardi \latin{et~al.}(1996)Bernardi, Olivucci, and
  Robb]{Bernardi_1996}
Bernardi,~F.; Olivucci,~M.; Robb,~M.~A. Potential Energy Surface Crossings in
  Organic Photochemistry. \emph{Chem. Soc. Rev.} \textbf{1996}, \emph{25},
  321\relax
\mciteBstWouldAddEndPuncttrue
\mciteSetBstMidEndSepPunct{\mcitedefaultmidpunct}
{\mcitedefaultendpunct}{\mcitedefaultseppunct}\relax
\EndOfBibitem
\bibitem[Olivucci(2010)]{Olivucci_2010}
Olivucci,~M. \emph{Computational Photochemistry}; {Elsevier Science}:
  Amsterdam; Boston (Mass.); Paris, 2010; OCLC: 800555856\relax
\mciteBstWouldAddEndPuncttrue
\mciteSetBstMidEndSepPunct{\mcitedefaultmidpunct}
{\mcitedefaultendpunct}{\mcitedefaultseppunct}\relax
\EndOfBibitem
\bibitem[Navizet \latin{et~al.}(2011)Navizet, Liu, Ferre, {Roca-Sanjun}, and
  Lindh]{Navizet_2011}
Navizet,~I.; Liu,~Y.-J.; Ferre,~N.; {Roca-Sanjun},~D.; Lindh,~R. The Chemistry
  of Bioluminescence: An Analysis of Chemical Functionalities.
  \emph{ChemPhysChem} \textbf{2011}, \emph{12}, 3064--3076\relax
\mciteBstWouldAddEndPuncttrue
\mciteSetBstMidEndSepPunct{\mcitedefaultmidpunct}
{\mcitedefaultendpunct}{\mcitedefaultseppunct}\relax
\EndOfBibitem
\bibitem[Robb \latin{et~al.}(2007)Robb, Garavelli, Olivucci, and
  Bernardi]{Robb_2007}
Robb,~M.~A.; Garavelli,~M.; Olivucci,~M.; Bernardi,~F. A {{Computational
  Strategy}} for {{Organic Photochemistry}}. In \emph{Reviews in
  {{Computational Chemistry}}}; Lipkowitz,~K.~B., Boyd,~D.~B., Eds.; {John
  Wiley \& Sons, Inc.}: Hoboken, NJ, USA, 2007; pp 87--146\relax
\mciteBstWouldAddEndPuncttrue
\mciteSetBstMidEndSepPunct{\mcitedefaultmidpunct}
{\mcitedefaultendpunct}{\mcitedefaultseppunct}\relax
\EndOfBibitem
\bibitem[Lazzeri \latin{et~al.}(2008)Lazzeri, Attaccalite, Wirtz, and
  Mauri]{Lazzeri_2008}
Lazzeri,~M.; Attaccalite,~C.; Wirtz,~L.; Mauri,~F. Impact of the
  Electron-Electron Correlation on Phonon Dispersion: Failure of LDA and GGA
  DFT Functionals in Graphene and Graphite. \emph{Phys. Rev. B} \textbf{2008},
  \emph{78}, 081406\relax
\mciteBstWouldAddEndPuncttrue
\mciteSetBstMidEndSepPunct{\mcitedefaultmidpunct}
{\mcitedefaultendpunct}{\mcitedefaultseppunct}\relax
\EndOfBibitem
\bibitem[Faber \latin{et~al.}(2011)Faber, Janssen, C\^ot\'e, Runge, and
  Blase]{Faber_2011b}
Faber,~C.; Janssen,~J.~L.; C\^ot\'e,~M.; Runge,~E.; Blase,~X. Electron-phonon
  coupling in the C${}_{60}$ fullerene within the many-body $GW$ approach.
  \emph{Phys. Rev. B} \textbf{2011}, \emph{84}, 155104\relax
\mciteBstWouldAddEndPuncttrue
\mciteSetBstMidEndSepPunct{\mcitedefaultmidpunct}
{\mcitedefaultendpunct}{\mcitedefaultseppunct}\relax
\EndOfBibitem
\bibitem[Yin \latin{et~al.}(2013)Yin, Kutepov, and Kotliar]{Yin_2013}
Yin,~Z.~P.; Kutepov,~A.; Kotliar,~G. Correlation-Enhanced Electron-Phonon
  Coupling: Applications of $GW$ and Screened Hybrid Functional to Bismuthates,
  Chloronitrides, and Other High-${T}_{c}$ Superconductors. \emph{Phys. Rev. X}
  \textbf{2013}, \emph{3}, 021011\relax
\mciteBstWouldAddEndPuncttrue
\mciteSetBstMidEndSepPunct{\mcitedefaultmidpunct}
{\mcitedefaultendpunct}{\mcitedefaultseppunct}\relax
\EndOfBibitem
\bibitem[Faber \latin{et~al.}(2015)Faber, Boulanger, Attaccalite, Cannuccia,
  Duchemin, Deutsch, and Blase]{Faber_2015}
Faber,~C.; Boulanger,~P.; Attaccalite,~C.; Cannuccia,~E.; Duchemin,~I.;
  Deutsch,~T.; Blase,~X. Exploring Approximations to the $GW$ Self-Energy Ionic
  Gradients. \emph{Phys. Rev. B} \textbf{2015}, \emph{91}, 155109\relax
\mciteBstWouldAddEndPuncttrue
\mciteSetBstMidEndSepPunct{\mcitedefaultmidpunct}
{\mcitedefaultendpunct}{\mcitedefaultseppunct}\relax
\EndOfBibitem
\bibitem[Monserrat(2016)]{Montserrat_2016}
Monserrat,~B. Correlation Effects on Electron-Phonon Coupling in
  Semiconductors: Many-Body Theory Along Thermal Lines. \emph{Phys. Rev. B}
  \textbf{2016}, \emph{93}, 100301\relax
\mciteBstWouldAddEndPuncttrue
\mciteSetBstMidEndSepPunct{\mcitedefaultmidpunct}
{\mcitedefaultendpunct}{\mcitedefaultseppunct}\relax
\EndOfBibitem
\bibitem[Li \latin{et~al.}(2019)Li, Antonius, Wu, da~Jornada, and
  Louie]{Zhenglu_2019}
Li,~Z.; Antonius,~G.; Wu,~M.; da~Jornada,~F.~H.; Louie,~S.~G. Electron-Phonon
  Coupling from Ab Initio Linear-Response Theory within the $GW$ Method:
  Correlation-Enhanced Interactions and Superconductivity in
  ${\mathrm{Ba}}_{1\ensuremath{-}x}{\mathrm{K}}_{x}{\mathrm{BiO}}_{3}$.
  \emph{Phys. Rev. Lett.} \textbf{2019}, \emph{122}, 186402\relax
\mciteBstWouldAddEndPuncttrue
\mciteSetBstMidEndSepPunct{\mcitedefaultmidpunct}
{\mcitedefaultendpunct}{\mcitedefaultseppunct}\relax
\EndOfBibitem
\bibitem[Ismail-Beigi and Louie(2003)Ismail-Beigi, and Louie]{Beigi_2003}
Ismail-Beigi,~S.; Louie,~S.~G. Excited-State Forces within a First-Principles
  Green's Function Formalism. \emph{Phys. Rev. Lett.} \textbf{2003}, \emph{90},
  076401\relax
\mciteBstWouldAddEndPuncttrue
\mciteSetBstMidEndSepPunct{\mcitedefaultmidpunct}
{\mcitedefaultendpunct}{\mcitedefaultseppunct}\relax
\EndOfBibitem
\bibitem[Hohenberg and Kohn(1964)Hohenberg, and Kohn]{Hohenberg_1964}
Hohenberg,~P.; Kohn,~W. Inhomogeneous Electron Gas. \emph{Phys. Rev.}
  \textbf{1964}, \emph{136}, B864--B871\relax
\mciteBstWouldAddEndPuncttrue
\mciteSetBstMidEndSepPunct{\mcitedefaultmidpunct}
{\mcitedefaultendpunct}{\mcitedefaultseppunct}\relax
\EndOfBibitem
\bibitem[Kohn and Sham(1965)Kohn, and Sham]{Kohn_1965}
Kohn,~W.; Sham,~L.~J. Self-Consistent Equations Including Exchange and
  Correlation Effects. \emph{Phys. Rev.} \textbf{1965}, \emph{140},
  A1133--A1138\relax
\mciteBstWouldAddEndPuncttrue
\mciteSetBstMidEndSepPunct{\mcitedefaultmidpunct}
{\mcitedefaultendpunct}{\mcitedefaultseppunct}\relax
\EndOfBibitem
\bibitem[Parr and Yang(1989)Parr, and Yang]{ParrBook}
Parr,~R.~G.; Yang,~W. \emph{Density-Functional Theory of Atoms and Molecules};
  Oxford: Clarendon Press, 1989\relax
\mciteBstWouldAddEndPuncttrue
\mciteSetBstMidEndSepPunct{\mcitedefaultmidpunct}
{\mcitedefaultendpunct}{\mcitedefaultseppunct}\relax
\EndOfBibitem
\bibitem[Olsen and Thygesen(2014)Olsen, and Thygesen]{Olsen_2014}
Olsen,~T.; Thygesen,~K.~S. Static Correlation Beyond the Random Phase
  Approximation: Dissociating H2 With the Bethe-Salpeter Equation and
  Time-Dependent GW. \emph{J. Chem. Phys.} \textbf{2014}, \emph{140},
  164116\relax
\mciteBstWouldAddEndPuncttrue
\mciteSetBstMidEndSepPunct{\mcitedefaultmidpunct}
{\mcitedefaultendpunct}{\mcitedefaultseppunct}\relax
\EndOfBibitem
\bibitem[Holzer \latin{et~al.}(2018)Holzer, Gui, Harding, Kresse, Helgaker, and
  Klopper]{Holzer_2018}
Holzer,~C.; Gui,~X.; Harding,~M.~E.; Kresse,~G.; Helgaker,~T.; Klopper,~W.
  Bethe--Salpeter Correlation Energies of Atoms and Molecules. \emph{J. Chem.
  Phys.} \textbf{2018}, \emph{149}, 144106\relax
\mciteBstWouldAddEndPuncttrue
\mciteSetBstMidEndSepPunct{\mcitedefaultmidpunct}
{\mcitedefaultendpunct}{\mcitedefaultseppunct}\relax
\EndOfBibitem
\bibitem[Li \latin{et~al.}(2019)Li, Drummond, Schuck, and Olevano]{Li_2019}
Li,~J.; Drummond,~N.~D.; Schuck,~P.; Olevano,~V. Comparing Many-Body Approaches
  Against the Helium Atom Exact Solution. \emph{SciPost Phys.} \textbf{2019},
  \emph{6}, 040\relax
\mciteBstWouldAddEndPuncttrue
\mciteSetBstMidEndSepPunct{\mcitedefaultmidpunct}
{\mcitedefaultendpunct}{\mcitedefaultseppunct}\relax
\EndOfBibitem
\bibitem[Li \latin{et~al.}(2020)Li, Duchemin, Blase, and Olevano]{Li_2020}
Li,~J.; Duchemin,~I.; Blase,~X.; Olevano,~V. Ground-State Correlation Energy of
  Beryllium Dimer by the Bethe--Salpeter Equation. \emph{SciPost Phys.}
  \textbf{2020}, \emph{8}, 020\relax
\mciteBstWouldAddEndPuncttrue
\mciteSetBstMidEndSepPunct{\mcitedefaultmidpunct}
{\mcitedefaultendpunct}{\mcitedefaultseppunct}\relax
\EndOfBibitem
\bibitem[Harding \latin{et~al.}(2008)Harding, Vazquez, Ruscic, Wilson, Gauss,
  and Stanton]{Harding_2008}
Harding,~M.~E.; Vazquez,~J.; Ruscic,~B.; Wilson,~A.~K.; Gauss,~J.;
  Stanton,~J.~F. High-Accuracy Extrapolated ab Initio Thermochemistry. III.
  Additional Improvements and Overview. \emph{J. Chem. Phys.} \textbf{2008},
  \emph{128}, 114111\relax
\mciteBstWouldAddEndPuncttrue
\mciteSetBstMidEndSepPunct{\mcitedefaultmidpunct}
{\mcitedefaultendpunct}{\mcitedefaultseppunct}\relax
\EndOfBibitem
\bibitem[Furche and {Van Voorhis}(2005)Furche, and {Van Voorhis}]{Furche_2005}
Furche,~F.; {Van Voorhis},~T. Fluctuation-Dissipation Theorem
  Density-Functional Theory. \emph{J. Chem. Phys.} \textbf{2005}, \emph{122},
  164106\relax
\mciteBstWouldAddEndPuncttrue
\mciteSetBstMidEndSepPunct{\mcitedefaultmidpunct}
{\mcitedefaultendpunct}{\mcitedefaultseppunct}\relax
\EndOfBibitem
\bibitem[Maggio and Kresse(2016)Maggio, and Kresse]{Maggio_2016}
Maggio,~E.; Kresse,~G. Correlation energy for the homogeneous electron gas:
  Exact Bethe-Salpeter solution and an approximate evaluation. \emph{Phys. Rev.
  B} \textbf{2016}, \emph{93}, 235113\relax
\mciteBstWouldAddEndPuncttrue
\mciteSetBstMidEndSepPunct{\mcitedefaultmidpunct}
{\mcitedefaultendpunct}{\mcitedefaultseppunct}\relax
\EndOfBibitem
\bibitem[Furche(2008)]{Furche_2008}
Furche,~F. Developing the Random Phase Approximation Into a Practical
  Post-{{Kohn}}\textendash{}{{Sham}} Correlation Model. \emph{J. Chem. Phys.}
  \textbf{2008}, \emph{129}, 114105\relax
\mciteBstWouldAddEndPuncttrue
\mciteSetBstMidEndSepPunct{\mcitedefaultmidpunct}
{\mcitedefaultendpunct}{\mcitedefaultseppunct}\relax
\EndOfBibitem
\bibitem[Toulouse \latin{et~al.}(2009)Toulouse, Gerber, Jansen, Savin, and
  Angyan]{Toulouse_2009}
Toulouse,~J.; Gerber,~I.~C.; Jansen,~G.; Savin,~A.; Angyan,~J.~G.
  Adiabatic-Connection Fluctuation-Dissipation Density-Functional Theory Based
  on Range Separation. \emph{Phys. Rev. Lett.} \textbf{2009}, \emph{102},
  096404\relax
\mciteBstWouldAddEndPuncttrue
\mciteSetBstMidEndSepPunct{\mcitedefaultmidpunct}
{\mcitedefaultendpunct}{\mcitedefaultseppunct}\relax
\EndOfBibitem
\bibitem[Toulouse \latin{et~al.}(2010)Toulouse, Zhu, Angyan, and
  Savin]{Toulouse_2010}
Toulouse,~J.; Zhu,~W.; Angyan,~J.~G.; Savin,~A. Range-Separated
  Density-Functional Theory With the Random-Phase Approximation: Detailed
  Formalism and Illustrative Applications. \emph{Phys. Rev. A} \textbf{2010},
  \emph{82}, 032502\relax
\mciteBstWouldAddEndPuncttrue
\mciteSetBstMidEndSepPunct{\mcitedefaultmidpunct}
{\mcitedefaultendpunct}{\mcitedefaultseppunct}\relax
\EndOfBibitem
\bibitem[Angyan \latin{et~al.}(2011)Angyan, Liu, Toulouse, and
  Jansen]{Angyan_2011}
Angyan,~J.~G.; Liu,~R.-F.; Toulouse,~J.; Jansen,~G. Correlation Energy
  Expressions from the Adiabatic-Connection Fluctuation Dissipation Theorem
  Approach. \emph{J. Chem. Theory Comput.} \textbf{2011}, \emph{7},
  3116--3130\relax
\mciteBstWouldAddEndPuncttrue
\mciteSetBstMidEndSepPunct{\mcitedefaultmidpunct}
{\mcitedefaultendpunct}{\mcitedefaultseppunct}\relax
\EndOfBibitem
\bibitem[Ren \latin{et~al.}(2012)Ren, Rinke, Joas, and Scheffler]{Ren_2012}
Ren,~X.; Rinke,~P.; Joas,~C.; Scheffler,~M. Random-Phase Approximation and Its
  Applications in Computational Chemistry and Materials Science. \emph{J.
  Mater. Sci.} \textbf{2012}, \emph{47}, 7447--7471\relax
\mciteBstWouldAddEndPuncttrue
\mciteSetBstMidEndSepPunct{\mcitedefaultmidpunct}
{\mcitedefaultendpunct}{\mcitedefaultseppunct}\relax
\EndOfBibitem
\bibitem[{van Setten} \latin{et~al.}(2015){van Setten}, Caruso, Sharifzadeh,
  Ren, Scheffler, Liu, Lischner, Lin, Deslippe, Louie, Yang, Weigend, Neaton,
  Evers, and Rinke]{vanSetten_2015}
{van Setten},~M.~J.; Caruso,~F.; Sharifzadeh,~S.; Ren,~X.; Scheffler,~M.;
  Liu,~F.; Lischner,~J.; Lin,~L.; Deslippe,~J.~R.; Louie,~S.~G.; Yang,~C.;
  Weigend,~F.; Neaton,~J.~B.; Evers,~F.; Rinke,~P. {{{\emph{GW}}}} 100:
  {{Benchmarking}}
  {{{\emph{G}}}}{\textsubscript{0}}{{{\emph{W}}}}{\textsubscript{0}} for
  {{Molecular Systems}}. \emph{J. Chem. Theory Comput.} \textbf{2015},
  \emph{11}, 5665--5687\relax
\mciteBstWouldAddEndPuncttrue
\mciteSetBstMidEndSepPunct{\mcitedefaultmidpunct}
{\mcitedefaultendpunct}{\mcitedefaultseppunct}\relax
\EndOfBibitem
\bibitem[Maggio \latin{et~al.}(2017)Maggio, Liu, {van Setten}, and
  Kresse]{Maggio_2017}
Maggio,~E.; Liu,~P.; {van Setten},~M.~J.; Kresse,~G. {{{\emph{GW}}}} 100: {{A
  Plane Wave Perspective}} for {{Small Molecules}}. \emph{J. Chem. Theory
  Comput.} \textbf{2017}, \emph{13}, 635--648\relax
\mciteBstWouldAddEndPuncttrue
\mciteSetBstMidEndSepPunct{\mcitedefaultmidpunct}
{\mcitedefaultendpunct}{\mcitedefaultseppunct}\relax
\EndOfBibitem
\bibitem[Loos \latin{et~al.}(2018)Loos, Romaniello, and Berger]{Loos_2018}
Loos,~P.~F.; Romaniello,~P.; Berger,~J.~A. Green Functions and
  Self-Consistency: Insights From the Spherium Model. \emph{J. Chem. Theory
  Comput.} \textbf{2018}, \emph{14}, 3071--3082\relax
\mciteBstWouldAddEndPuncttrue
\mciteSetBstMidEndSepPunct{\mcitedefaultmidpunct}
{\mcitedefaultendpunct}{\mcitedefaultseppunct}\relax
\EndOfBibitem
\bibitem[Veril \latin{et~al.}(2018)Veril, Romaniello, Berger, and
  Loos]{Veril_2018}
Veril,~M.; Romaniello,~P.; Berger,~J.~A.; Loos,~P.~F. Unphysical
  Discontinuities in GW Methods. \emph{J. Chem. Theory Comput.} \textbf{2018},
  \emph{14}, 5220\relax
\mciteBstWouldAddEndPuncttrue
\mciteSetBstMidEndSepPunct{\mcitedefaultmidpunct}
{\mcitedefaultendpunct}{\mcitedefaultseppunct}\relax
\EndOfBibitem
\bibitem[Duchemin and Blase(2020)Duchemin, and Blase]{Duchemin_2020}
Duchemin,~I.; Blase,~X. Robust Analytic Continuation Approach to Many-Body GW
  Calculations. \emph{J. Chem. Theory Comput (accepted)} \textbf{2020}, \relax
\mciteBstWouldAddEndPunctfalse
\mciteSetBstMidEndSepPunct{\mcitedefaultmidpunct}
{}{\mcitedefaultseppunct}\relax
\EndOfBibitem
\bibitem[Martin \latin{et~al.}(2016)Martin, Reining, and Ceperley]{Martin_2016}
Martin,~R.~M.; Reining,~L.; Ceperley,~D.~M. \emph{Interacting Electrons: Theory
  and Computational Approaches}; Cambridge University Press, 2016\relax
\mciteBstWouldAddEndPuncttrue
\mciteSetBstMidEndSepPunct{\mcitedefaultmidpunct}
{\mcitedefaultendpunct}{\mcitedefaultseppunct}\relax
\EndOfBibitem
\bibitem[Hedin(1965)]{Hedin_1965}
Hedin,~L. New Method for Calculating the One-Particle {{Green}}'s Function with
  Application to the Electron-Gas Problem. \emph{Phys. Rev.} \textbf{1965},
  \emph{139}, A796\relax
\mciteBstWouldAddEndPuncttrue
\mciteSetBstMidEndSepPunct{\mcitedefaultmidpunct}
{\mcitedefaultendpunct}{\mcitedefaultseppunct}\relax
\EndOfBibitem
\bibitem[Aryasetiawan and Gunnarsson(1998)Aryasetiawan, and
  Gunnarsson]{Aryasetiawan_1998}
Aryasetiawan,~F.; Gunnarsson,~O. The GW Method. \emph{Rep. Prog. Phys.}
  \textbf{1998}, \emph{61}, 237--312\relax
\mciteBstWouldAddEndPuncttrue
\mciteSetBstMidEndSepPunct{\mcitedefaultmidpunct}
{\mcitedefaultendpunct}{\mcitedefaultseppunct}\relax
\EndOfBibitem
\bibitem[Onida \latin{et~al.}(2002)Onida, Reining, and and]{Onida_2002}
Onida,~G.; Reining,~L.; and,~A.~R. Electronic Excitations: Density-Functional
  Versus Many-Body Green's Function Approaches. \emph{Rev. Mod. Phys.}
  \textbf{2002}, \emph{74}, 601--659\relax
\mciteBstWouldAddEndPuncttrue
\mciteSetBstMidEndSepPunct{\mcitedefaultmidpunct}
{\mcitedefaultendpunct}{\mcitedefaultseppunct}\relax
\EndOfBibitem
\bibitem[Reining(2017)]{Reining_2017}
Reining,~L. The {{GW}} Approximation: Content, Successes and Limitations: {{The
  GW}} Approximation. \emph{Wiley Interdiscip. Rev. Comput. Mol. Sci.}
  \textbf{2017}, e1344\relax
\mciteBstWouldAddEndPuncttrue
\mciteSetBstMidEndSepPunct{\mcitedefaultmidpunct}
{\mcitedefaultendpunct}{\mcitedefaultseppunct}\relax
\EndOfBibitem
\bibitem[Hanke and Sham(1980)Hanke, and Sham]{Hanke_1980}
Hanke,~W.; Sham,~L.~J. Many-Particle Effects in the Optical Spectrum of a
  Semiconductor. \emph{Phys. Rev. B} \textbf{1980}, \emph{21}, 4656\relax
\mciteBstWouldAddEndPuncttrue
\mciteSetBstMidEndSepPunct{\mcitedefaultmidpunct}
{\mcitedefaultendpunct}{\mcitedefaultseppunct}\relax
\EndOfBibitem
\bibitem[Strinati(1984)]{Strinati_1984}
Strinati,~G. Effects of Dynamical Screening on Resonances at Inner-Shell
  Thresholds in Semiconductors. \emph{Phys. Rev. B} \textbf{1984}, \emph{29},
  5718\relax
\mciteBstWouldAddEndPuncttrue
\mciteSetBstMidEndSepPunct{\mcitedefaultmidpunct}
{\mcitedefaultendpunct}{\mcitedefaultseppunct}\relax
\EndOfBibitem
\bibitem[Strinati(1982)]{Strinati_1982}
Strinati,~G. Dynamical Shift and Broadening of Core Excitons in Semiconductors.
  \emph{Phys. Rev. Lett.} \textbf{1982}, \emph{49}, 1519\relax
\mciteBstWouldAddEndPuncttrue
\mciteSetBstMidEndSepPunct{\mcitedefaultmidpunct}
{\mcitedefaultendpunct}{\mcitedefaultseppunct}\relax
\EndOfBibitem
\bibitem[Dreuw and Head-Gordon(2005)Dreuw, and Head-Gordon]{Dreuw_2005}
Dreuw,~A.; Head-Gordon,~M. Single-{{Reference}} Ab {{Initio Methods}} for the
  {{Calculation}} of {{Excited States}} of {{Large Molecules}}. \emph{Chem.
  Rev.} \textbf{2005}, \emph{105}, 4009--4037\relax
\mciteBstWouldAddEndPuncttrue
\mciteSetBstMidEndSepPunct{\mcitedefaultmidpunct}
{\mcitedefaultendpunct}{\mcitedefaultseppunct}\relax
\EndOfBibitem
\bibitem[Holzer \latin{et~al.}(2019)Holzer, Teale, Hampe, Stopkowicz, Helgaker,
  and Klopper]{Holzer_2019}
Holzer,~C.; Teale,~A.~M.; Hampe,~F.; Stopkowicz,~S.; Helgaker,~T.; Klopper,~W.
  GW Quasiparticle Energies of Atoms in Strong Magnetic Fields. \emph{J. Chem.
  Phys.} \textbf{2019}, \emph{150}, 214112\relax
\mciteBstWouldAddEndPuncttrue
\mciteSetBstMidEndSepPunct{\mcitedefaultmidpunct}
{\mcitedefaultendpunct}{\mcitedefaultseppunct}\relax
\EndOfBibitem
\bibitem[Colonna \latin{et~al.}(2014)Colonna, Hellgren, and {de
  Gironcoli}]{Colonna_2014}
Colonna,~N.; Hellgren,~M.; {de Gironcoli},~S. Correlation Energy Within
  Exact-Exchange Adiabatic Connection Fluctuation-Dissipation Theory:
  Systematic Development and Simple Approximations. \emph{Phys. Rev. B}
  \textbf{2014}, \emph{90}, 125150\relax
\mciteBstWouldAddEndPuncttrue
\mciteSetBstMidEndSepPunct{\mcitedefaultmidpunct}
{\mcitedefaultendpunct}{\mcitedefaultseppunct}\relax
\EndOfBibitem
\bibitem[Hybertsen and Louie(1985)Hybertsen, and Louie]{Hybertsen_1985a}
Hybertsen,~M.~S.; Louie,~S.~G. First-{{Principles Theory}} of
  {{Quasiparticles}}: {{Calculation}} of {{Band Gaps}} in {{Semiconductors}}
  and {{Insulators}}. \emph{Phys. Rev. Lett.} \textbf{1985}, \emph{55},
  1418--1421\relax
\mciteBstWouldAddEndPuncttrue
\mciteSetBstMidEndSepPunct{\mcitedefaultmidpunct}
{\mcitedefaultendpunct}{\mcitedefaultseppunct}\relax
\EndOfBibitem
\bibitem[Christiansen \latin{et~al.}(1995)Christiansen, Koch, and
  J{\o}rgensen]{Christiansen_1995}
Christiansen,~O.; Koch,~H.; J{\o}rgensen,~P. The Second-Order Approximate
  Coupled Cluster Singles and Doubles Model CC2. \emph{Chem. Phys. Lett.}
  \textbf{1995}, \emph{243}, 409--418\relax
\mciteBstWouldAddEndPuncttrue
\mciteSetBstMidEndSepPunct{\mcitedefaultmidpunct}
{\mcitedefaultendpunct}{\mcitedefaultseppunct}\relax
\EndOfBibitem
\bibitem[Purvis~III and Bartlett(1982)Purvis~III, and Bartlett]{Purvis_1982}
Purvis~III,~G.~P.; Bartlett,~R.~J. A Full Coupled-Cluster Singles and Doubles
  Model: The Inclusion of Disconnected Triples. \emph{J. Chem. Phys.}
  \textbf{1982}, \emph{76}, 1910--1918\relax
\mciteBstWouldAddEndPuncttrue
\mciteSetBstMidEndSepPunct{\mcitedefaultmidpunct}
{\mcitedefaultendpunct}{\mcitedefaultseppunct}\relax
\EndOfBibitem
\bibitem[Christiansen \latin{et~al.}(1995)Christiansen, Koch, and
  J{\o}rgensen]{Christiansen_1995b}
Christiansen,~O.; Koch,~H.; J{\o}rgensen,~P. Response Functions in the CC3
  Iterative Triple Excitation Model. \emph{J. Chem. Phys.} \textbf{1995},
  \emph{103}, 7429--7441\relax
\mciteBstWouldAddEndPuncttrue
\mciteSetBstMidEndSepPunct{\mcitedefaultmidpunct}
{\mcitedefaultendpunct}{\mcitedefaultseppunct}\relax
\EndOfBibitem
\bibitem[Aidas \latin{et~al.}(2014)Aidas, Angeli, Bak, Bakken, Bast, Boman,
  Christiansen, Cimiraglia, Coriani, Dahle, Dalskov, Ekstr{\"o}m, Enevoldsen,
  Eriksen, Ettenhuber, Fern{\'a}ndez, Ferrighi, Fliegl, Frediani, Hald,
  Halkier, H{\"a}ttig, Heiberg, Helgaker, Hennum, Hettema, Hjerten{\ae}s,
  H{\o}st, H{\o}yvik, Iozzi, Jans{\'\i}k, Jensen, Jonsson, J{\o}rgensen,
  Kauczor, Kirpekar, Kj{\ae}rgaard, Klopper, Knecht, Kobayashi, Koch, Kongsted,
  Krapp, Kristensen, Ligabue, Lutn{\ae}s, Melo, Mikkelsen, Myhre, Neiss,
  Nielsen, Norman, Olsen, Olsen, Osted, Packer, Pawlowski, Pedersen, Provasi,
  Reine, Rinkevicius, Ruden, Ruud, Rybkin, Sa{\l}ek, Samson, de~Mer{\'a}s,
  Saue, Sauer, Schimmelpfennig, Sneskov, Steindal, Sylvester-Hvid, Taylor,
  Teale, Tellgren, Tew, Thorvaldsen, Th{\o}gersen, Vahtras, Watson, Wilson,
  Ziolkowski, and {\AA}gren]{dalton}
Aidas,~K. \latin{et~al.}  The Dalton Quantum Chemistry Program System.
  \emph{WIREs Comput. Mol. Sci.} \textbf{2014}, \emph{4}, 269--284\relax
\mciteBstWouldAddEndPuncttrue
\mciteSetBstMidEndSepPunct{\mcitedefaultmidpunct}
{\mcitedefaultendpunct}{\mcitedefaultseppunct}\relax
\EndOfBibitem
\bibitem[Parrish \latin{et~al.}(2017)Parrish, Burns, Smith, Simmonett,
  DePrince, Hohenstein, Bozkaya, Sokolov, Di~Remigio, Richard, Gonthier, James,
  McAlexander, Kumar, Saitow, Wang, Pritchard, Verma, Schaefer, Patkowski,
  King, Valeev, Evangelista, Turney, Crawford, and Sherrill]{Psi4}
Parrish,~R.~M.; Burns,~L.~A.; Smith,~D. G.~A.; Simmonett,~A.~C.;
  DePrince,~A.~E.; Hohenstein,~E.~G.; Bozkaya,~U.; Sokolov,~A.~Y.;
  Di~Remigio,~R.; Richard,~R.~M.; Gonthier,~J.~F.; James,~A.~M.;
  McAlexander,~H.~R.; Kumar,~A.; Saitow,~M.; Wang,~X.; Pritchard,~B.~P.;
  Verma,~P.; Schaefer,~H.~F.; Patkowski,~K.; King,~R.~A.; Valeev,~E.~F.;
  Evangelista,~F.~A.; Turney,~J.~M.; Crawford,~T.~D.; Sherrill,~C.~D. Psi4 1.1:
  An Open-Source Electronic Structure Program Emphasizing Automation, Advanced
  Libraries, and Interoperability. \emph{J. Chem. Theory Comput.}
  \textbf{2017}, \emph{13}, 3185--3197, PMID: 28489372\relax
\mciteBstWouldAddEndPuncttrue
\mciteSetBstMidEndSepPunct{\mcitedefaultmidpunct}
{\mcitedefaultendpunct}{\mcitedefaultseppunct}\relax
\EndOfBibitem
\bibitem[H{\"a}ttig(2005)]{Hattig_2005c}
H{\"a}ttig,~C. Structure Optimizations for Excited States with Correlated
  Second-Order Methods: CC2 and ADC(2). In \emph{Response Theory and Molecular
  Properties (A Tribute to Jan Linderberg and Poul J{\o}rgensen)};
  Jensen,~H.~A., Ed.; Advances in Quantum Chemistry; Academic Press, 2005;
  Vol.~50; pp 37--60\relax
\mciteBstWouldAddEndPuncttrue
\mciteSetBstMidEndSepPunct{\mcitedefaultmidpunct}
{\mcitedefaultendpunct}{\mcitedefaultseppunct}\relax
\EndOfBibitem
\bibitem[Budz{\'a}k \latin{et~al.}(2017)Budz{\'a}k, Scalmani, and
  Jacquemin]{Budzak_2017}
Budz{\'a}k,~{\v S}.; Scalmani,~G.; Jacquemin,~D. Accurate Excited-State
  Geometries: a CASPT2 and Coupled-Cluster Reference Database for Small
  Molecules. \emph{J. Chem. Theory Comput.} \textbf{2017}, \emph{13},
  6237--6252\relax
\mciteBstWouldAddEndPuncttrue
\mciteSetBstMidEndSepPunct{\mcitedefaultmidpunct}
{\mcitedefaultendpunct}{\mcitedefaultseppunct}\relax
\EndOfBibitem
\bibitem[K{\'a}llay \latin{et~al.}(2017)K{\'a}llay, Rolik, Csontos, Nagy, Samu,
  Mester, Cs{\'o}ka, Szab{\'o}, Ladj{\'a}nszki, Szegedy, Lad{\'o}czki, Petrov,
  Farkas, Mezei, and H{\'e}gely.]{MRCC}
K{\'a}llay,~M.; Rolik,~Z.; Csontos,~J.; Nagy,~P.; Samu,~G.; Mester,~D.;
  Cs{\'o}ka,~J.; Szab{\'o},~B.; Ladj{\'a}nszki,~I.; Szegedy,~L.;
  Lad{\'o}czki,~B.; Petrov,~K.; Farkas,~M.; Mezei,~P.~D.; H{\'e}gely.,~B. MRCC,
  Quantum Chemical Program. 2017\relax
\mciteBstWouldAddEndPuncttrue
\mciteSetBstMidEndSepPunct{\mcitedefaultmidpunct}
{\mcitedefaultendpunct}{\mcitedefaultseppunct}\relax
\EndOfBibitem
\bibitem[Duchemin and Blase(2019)Duchemin, and Blase]{Duchemin_2019}
Duchemin,~I.; Blase,~X. Separable Resolution-of-the-Identity with All-Electron
  Gaussian Bases: Application to Cubic-scaling RPA. \emph{J. Chem. Phys.}
  \textbf{2019}, \emph{150}, 174120\relax
\mciteBstWouldAddEndPuncttrue
\mciteSetBstMidEndSepPunct{\mcitedefaultmidpunct}
{\mcitedefaultendpunct}{\mcitedefaultseppunct}\relax
\EndOfBibitem
\bibitem[Garniron \latin{et~al.}(2019)Garniron, Gasperich, Applencourt, Benali,
  Fert{\'e}, Paquier, Pradines, Assaraf, Reinhardt, Toulouse, Barbaresco,
  Renon, David, Malrieu, V{\'e}ril, Caffarel, Loos, Giner, and Scemama]{QP2}
Garniron,~Y.; Gasperich,~K.; Applencourt,~T.; Benali,~A.; Fert{\'e},~A.;
  Paquier,~J.; Pradines,~B.; Assaraf,~R.; Reinhardt,~P.; Toulouse,~J.;
  Barbaresco,~P.; Renon,~N.; David,~G.; Malrieu,~J.~P.; V{\'e}ril,~M.;
  Caffarel,~M.; Loos,~P.~F.; Giner,~E.; Scemama,~A. Quantum Package 2.0: A
  Open-Source Determinant-Driven Suite Of Programs. \emph{J. Chem. Theory
  Comput.} \textbf{2019}, \emph{15}, 3591\relax
\mciteBstWouldAddEndPuncttrue
\mciteSetBstMidEndSepPunct{\mcitedefaultmidpunct}
{\mcitedefaultendpunct}{\mcitedefaultseppunct}\relax
\EndOfBibitem
\bibitem[Caruso \latin{et~al.}(2013)Caruso, Rohr, Hellgren, Ren, Rinke, Rubio,
  and Scheffler]{Caruso_2013}
Caruso,~F.; Rohr,~D.~R.; Hellgren,~M.; Ren,~X.; Rinke,~P.; Rubio,~A.;
  Scheffler,~M. Bond {{Breaking}} and {{Bond Formation}}: {{How Electron
  Correlation}} Is {{Captured}} in {{Many}}-{{Body Perturbation Theory}} and
  {{Density}}-{{Functional Theory}}. \emph{Phys. Rev. Lett.} \textbf{2013},
  \emph{110}, 146403\relax
\mciteBstWouldAddEndPuncttrue
\mciteSetBstMidEndSepPunct{\mcitedefaultmidpunct}
{\mcitedefaultendpunct}{\mcitedefaultseppunct}\relax
\EndOfBibitem
\bibitem[Hellgren \latin{et~al.}(2015)Hellgren, Caruso, Rohr, Ren, Rubio,
  Scheffler, and Rinke]{Hellgren_2015}
Hellgren,~M.; Caruso,~F.; Rohr,~D.~R.; Ren,~X.; Rubio,~A.; Scheffler,~M.;
  Rinke,~P. Static Correlation and Electron Localization in Molecular Dimers
  from the Self-Consistent {{RPA}} and {{G W}} Approximation. \emph{Phys. Rev.
  B} \textbf{2015}, \emph{91}, 165110\relax
\mciteBstWouldAddEndPuncttrue
\mciteSetBstMidEndSepPunct{\mcitedefaultmidpunct}
{\mcitedefaultendpunct}{\mcitedefaultseppunct}\relax
\EndOfBibitem
\end{mcitethebibliography}

\end{document}